\documentstyle[11pt,aaspp]{article}
\input epsf

\def\gtorder{\mathrel{\raise.3ex\hbox{$>$}\mkern-14mu
             \lower0.6ex\hbox{$\sim$}}}

\def\ltorder{\mathrel{\raise.3ex\hbox{$<$}\mkern-14mu
             \lower0.6ex\hbox{$\sim$}}}

\def\simless{\mathbin{\lower 3pt\hbox
   {$\rlap{\raise 5pt\hbox{$\char'074$}}\mathchar"7218$}}} %< or of order
\def\simgreat{\mathbin{\lower 3pt\hbox
   {$\rlap{\raise 5pt\hbox{$\char'076$}}\mathchar"7218$}}} %> or of order
\def\be{\begin{equation}}
\def\ee{\end{equation}}

\input{postscript.sty}
\postscripton

\newcommand{\half}{{\textstyle{1\over2}}}
\newcommand{\absa}{\beta}
\newcommand{\ylm}{{\rm Y}_{\ell m}(\theta,\phi)}
\begin{document}

\title{A map for eccentric orbits in triaxial potentials}
 
\author{Jihad Touma$^{1,3}$ and Scott Tremaine$^{2,3}$}
 
\affil{$^1$McDonald Observatory, University of Texas, RLM 16.228, Austin, TX
78712, USA}
\affil{$^2$Canadian Institute for Advanced Research, Program in Cosmology and
Gravity}
\affil{$^3$Canadian Institute for Theoretical Astrophysics, McLennan Labs, 
University of Toronto,\\ 60~St.\ George St., Toronto M5S 3H8, Canada}
 
\begin{abstract} 

We construct a simple symplectic map to study the dynamics of eccentric orbits
in non-spherical potentials. The map offers a dramatic improvement in speed
over traditional integration methods, while accurately representing the
qualitative details of the dynamics. We focus attention on planar,
non-axisymmetric power-law potentials, in particular the logarithmic
potential. We confirm the presence of resonant orbit families (``boxlets'') in
this potential and uncover new dynamics such as the emergence of a stochastic
web in nearly axisymmetric logarithmic potentials. The map can also be applied
to triaxial, lopsided, non-power-law and rotating potentials.

\end{abstract}
 
\section{Introduction}

The morphology of orbits in triaxial potentials determines the structure of
triaxial galaxies. The simplest triaxial potentials are those of St\"ackel
form, which support four major families of regular orbits (boxes, short-axis
tubes, and inner and outer long-axis tubes) and no stochastic orbits
(Lynden-Bell 1962, de Zeeuw
1985). Many plausible triaxial potentials with smooth cores exhibit similar
structure: most of phase space is occupied by these four major orbit families,
with the leftover phase space hosting resonant minor families and stochastic
orbits (Schwarzschild 1979).

Potentials without cores, such as triaxial logarithmic potentials of the form
$\ln(\sum x_i^2/a_i^2)$, can have quite different behavior (Gerhard \& Binney
1985, Miralda-Escud\'e \& Schwarzschild 1989). In this case the box orbits are
replaced by a number of minor orbit families corresponding to various
resonances. The minor orbits are centrophobic (center-avoiding) in contrast to
the centrophilic box orbits, which all pass arbitrarily close to the center.

In special cases non-axisymmetric power-law potentials can have
St\"ackel form. Sridhar \& Touma (1997) constructed planar,
non-axisymmetric cuspy potentials which are separable in parabolic
coordinates. These potentials remain separable if a central black hole
is added.  In the absence of a black hole, the dynamics are scale-free
and all orbits belong to the centrophobic
``banana'' family. With a black hole, an additional family of
centrophilic orbits (``lenses'') emerges to replace the box orbits of
models with a smooth core.

The relevance of power-law potentials is enhanced by recent
observations. High-resolution {\it Hubble Space Telescope} photometry
of nearby elliptical galaxies and spiral bulges shows that the
luminosity density near the center is a power-law, $\nu(r)\propto
r^{\alpha-2}$ with $-0.3\simless \alpha\simless 1.7$ (e.g. Gebhardt et
al. 1996). Few or no galaxies have smooth cores ($\alpha=2$). There is
also growing evidence for massive black holes in the centers of many
nearby galaxies (Kormendy \& Richstone 1995), which generate singular
potentials that share many properties with the potentials generated by
power-law densities.

We would like to understand how the orbit structure in non-axisymmetric
power-law potentials $\propto r^\alpha$ changes with the exponent $\alpha$,
and with the degree of non-axisymmetry. An exhaustive study with conventional
integration methods is costly and difficult to interpret. We show how the
dynamics of eccentric orbits in power-law potentials can be illuminated with
the help of a symplectic mapping. The mapping models the evolution of such
orbits as a two-step process: (i) precession of the orientation of the orbit
in an axisymmetric potential; (ii) a kick to the angular momentum of the
orbits at apocentre (where the star spends most of its time, and the torques
are likely to be strongest). We use this mapping to study the orbital
structure of non-axisymmetric power-law potentials over the range of
$\alpha$ relevant to galaxies.

\subsection{Scale-free spherical potentials}

We assemble some properties of orbits in attracting spherical
potentials of the form 
\be
\Phi_\alpha(r)=\left\{\begin{array}{cc}Kr^\alpha, &\alpha\not=0, \\
K\ln r &\alpha=0. \end{array} \right.
\label{eq:potdef}
\ee 
where we assume henceforth that $-1\le\alpha\le 2$, which is true for most
plausible potentials ($\alpha=-1$ is Keplerian; $\alpha=2$ is the harmonic
oscillator; $\alpha=0$ is the singular isothermal sphere). Potentials with
smaller $\alpha$ arise from density distributions with greater central
concentration. We shall call potentials with $\alpha>0$ concave potentials and
those with $\alpha<0$ convex.  With no loss of generality we can set
$K=\hbox{sgn}(\alpha)$ for $\alpha\not=0$ and $K=1$ for $\alpha=0$.  The
energy corresponding to a circular orbit with radius $r_c$ is then
\be
E=\left\{\begin{array}{cc}\half|\alpha|r_c^\alpha+\hbox{sgn}(\alpha)r_c^\alpha,
&\alpha\not=0, \\ \half+\ln r_c, &\alpha=0. \end{array} \right.  
\ee 
Note that
$\hbox{sgn}(E)=\hbox{sgn}(\alpha)$ for 
all non-zero $\alpha>-2$, so that \be
|E|=\left(1+\half\alpha\right) r_c^\alpha,\qquad \alpha\not=0.  
\ee 
The angular momentum of a circular orbit is 
\be
L_c(E)=\left\{\begin{array}{cc}|\alpha|^{1/2}
\left(1+\half\alpha\right)^{-{\textstyle {1\over\alpha}}-\half}
|E|^{{\textstyle {1\over\alpha}}+\half} \equiv h(\alpha)|E|^{{\textstyle
{1\over\alpha}}+\half}, &\alpha\not=0,\\ \exp\left(-\half\right)\exp(E)\equiv
h(0)\exp(E),&\alpha=0.
\end{array} \right.
\label{eq:ellmax}
\ee 
Now consider motion in a plane and let $L$ be the scalar angular momentum,
positive for prograde orbits and negative for retrograde orbits; it is
convenient to work with the dimensionless angular momentum 
\be 
y\equiv {L\over L_c(E)},
\label{eq:ydef}
\ee
which can vary from $-1$ to $+1$. The orientation of an eccentric orbit
can be specified by the azimuthal angle of its $n^{\rm th}$ apocenter,
$\phi_n$. We may write
\be
\phi_{n+1}=\phi_n+g(\alpha,y);
\ee
where the precession rate $g$ is independent of energy because the potential
is scale-free, and $|g(\alpha,y)|$ is $2\pi P_r/P_\phi$, where $P_r$ and
$P_\phi$ are the radial and azimuthal periods. The function $g$ is odd in $y$,
and is given by 
\be
g(\alpha,y)=2h(\alpha)y\left\{\begin{array}{cc}\displaystyle
\int {\displaystyle dr\over
\displaystyle r[2\,\hbox{sgn}(\alpha)(r^2-r^{\alpha+2})-y^2h^2(\alpha)]^{1/2}},
&\alpha\not=0,\\ \displaystyle
\int{\displaystyle dr\over\displaystyle r[-2r^2\ln
r-y^2h^2(0)]^{1/2}}, &\alpha=0, \end{array}\right.
\label{eq:gdef}
\ee
where the integral is over all radii for which the radicand is positive. 
For near-radial orbits we have
\be
\lim_{y\to 0\pm}g(\alpha,y)\equiv \pm g_0(\alpha),
\label{eq:gzero}
\ee
where 
\be
g_0(\alpha)=\cases{\pi,&for $\alpha\ge0$,\cr
                   {\displaystyle 2\pi\over\displaystyle  2+\alpha},
                      &for $\alpha<0$.\cr}
\label{eq:zerolim}
\ee
When $|y|$ is small we have (cf. Appendix A)
\be
g(\alpha,y)-g_0(\alpha)\sim |y|^\beta,
\label{eq:gasymp}
\ee
where
\be
\beta(\alpha)=\left\{\begin{array}{cc}1&1\le\alpha<2,\\
               \alpha&0<\alpha\le1,\\
               -{\displaystyle 2\alpha\over \displaystyle 
                    2+\alpha}&-\frac{2}{3}\le\alpha<0,\\
	       1&-1<\alpha\le-\frac{2}{3}.\end{array}\right.
\label{eq:ggasymp}
\ee
For near-circular orbits we have
\be
\lim_{y\to \pm1}g(\alpha,y)=\pm {2\pi\over(2+\alpha)^{1/2}},
\ee
which is the usual epicyclic approximation. There are two special cases for
which the precession rate is simple, 
\be
g(\alpha,y)=\hbox{sgn}(y)\left\{\begin{array}{cc}\pi,&\alpha=2,\\
                                            2\pi,&\alpha=-1;\end{array}\right.
\label{eq:spec}
\ee
in addition, for $\alpha={2\over3},1,-{1\over2},-{2\over3}$ the precession
rate can be expressed in terms of elliptic functions 
(Whittaker 1959). In general
$g(\alpha,y)$ cannot be determined in closed form, although an asymptotic
series is given in Appendix A. A useful identity, easily derived
from (\ref{eq:gdef}), is (Grant \& Rosner 1994)
\be
g(\overline{\alpha},y)=\left(1+\half\alpha\right)g(\alpha,y)\qquad\hbox{where}
\qquad \alpha=-{2\overline{\alpha}\over2+\overline{\alpha}};
\label{eq:dual}
\ee
thus the behavior of $g$ in the range $-1\le\alpha<0$ is determined by its
behavior in the range $0<\alpha\le 2$.

\begin{figure}
\centerline{\hbox{\epsfxsize=5in\epsfbox{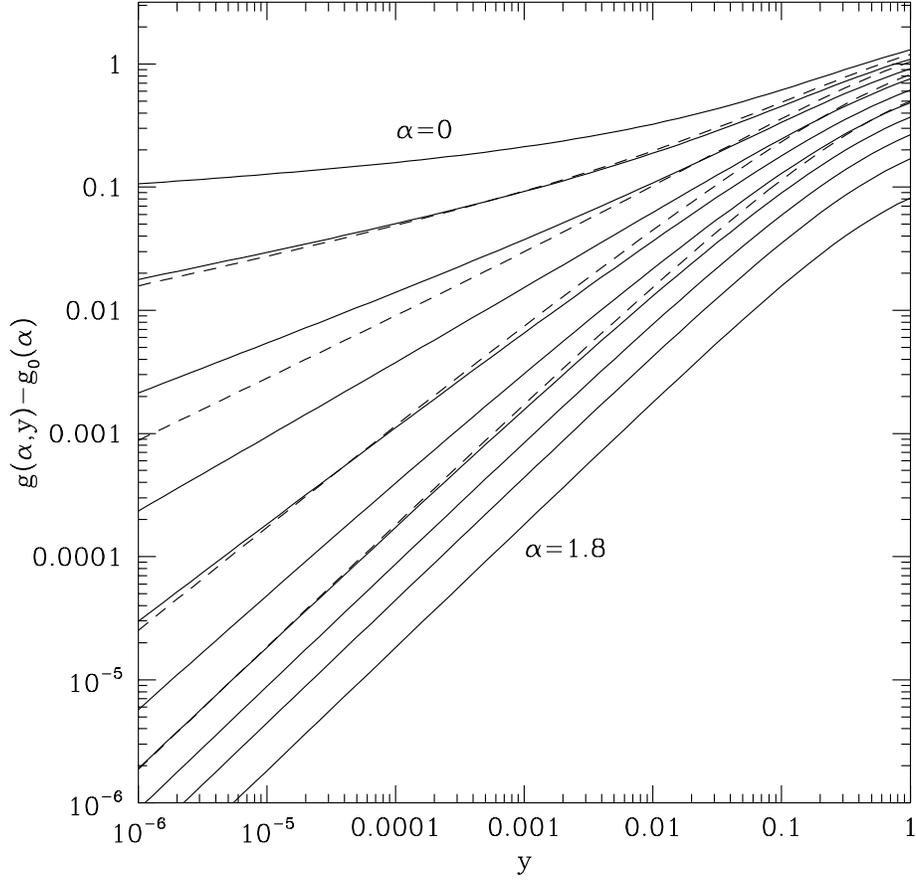}}}
\caption[Figure 1]{Precession rate $g(\alpha,y)$ (eq. \ref{eq:gdef}) as a
function of dimensionless angular momentum $y$, for power-law potentials with
exponent $\alpha=-0.8(0.2)1.8$. The precession rate is an odd function of $y$
and is only shown for $y>0$. Potentials with $\alpha<0$ are represented by
dashed lines, those with $\alpha\ge0$ by solid lines. The value of $g$ at
$y=0$ (eq. \ref{eq:zerolim}) has been subtracted off; thus the analogous curves
for $\alpha=-1$ and $2$ are zero everywhere (cf. eq. \ref{eq:spec}).}
\label{fi:g}
\end{figure}

The map we discuss in the following section repeatedly employs the function
$g(\alpha,y)$ at fixed $\alpha$. To minimize calculations of the integral
(\ref{eq:gdef}) we tabulate $\log_{10}[g(\alpha,y)-g_0(\alpha)]$ at
$\log_{10}(y)=-6(0.1)0$ and interpolate using a cubic spline. Computing this
integral is straightforward but difficult to do well, in part because of the
integrable singularities at the endpoints and in part because the endpoints
themselves are determined implicitly; we can supply our program or results upon
request. Figure \ref{fi:g} shows $g(\alpha,y)-g_0(\alpha)$. 

\section{A map for eccentric orbits}\label{sec:mapdef}

We now examine the behaviour of orbits in a scale-free axisymmetric potential
that is perturbed by a small non-axisymmetric potential. If the unperturbed
orbit is nearly circular, the behaviour can be analyzed using the epicycle
approximation (e.g. Binney \& Tremaine 1987, \S3.3.3). Here
we focus instead on an approximation valid for eccentric orbits.

In most cases the torques exerted by a non-axisymmetric potential on an
eccentric orbit are concentrated near apocenter, since (i) the lever arm is
larger; (ii) the particle spends most of its time near apocenter; (iii) any
external tidal forces are stronger at larger radii.

Let $y_n$ and $\phi_n$ be the values of the dimensionless angular momentum and
the azimuthal angle of a particle at its $n^{\rm th}$ apocenter passage. If
the non-axisymmetric potential has $\exp(im\phi)$ symmetry then the
time-integrated torque over the $n^{\rm th}$ apocenter passage can be written
$-\epsilon L_c(E)\sin m\phi_n$; the minus sign is appropriate if the azimuthal
minimum of the non-axisymmetric potential lies along the ray $\phi=0$.  This
torque induces a change in the angular momentum, of which half occurs
before apocenter and half after; thus the dimensionless angular
momentum after the particle leaves apocenter is 
\be 
y_n'=y_n-\half\epsilon\sin m\phi_n.
\label{eq:mapone}
\ee
The position of the following apocenter is 
\be
\phi_{n+1}=\phi_n+g(\alpha,y_n'),
\label{eq:maptwo}
\ee
and the angular momentum at this apocenter is
\be
y_{n+1}=y_n'-\half\epsilon\sin m\phi_{n+1}.
\label{eq:mapthree}
\ee

Equations (\ref{eq:mapone}--\ref{eq:mapthree}) define a simple map
$(\phi_n,y_n)\to (\phi_{n+1},y_{n+1})$ that describes the dynamics of
eccentric orbits. The map is symplectic, like the exact equations of motion,
and is more general than the exact equations because it does not depend on the
specific radial form of the non-axisymmetric potential, so long as the torque
is concentrated near apocenter. 

\subsection{Relation to the standard map}\label{sec:std}

When $1<\alpha<2$ equations (\ref{eq:gzero})--(\ref{eq:ggasymp}) imply
that $g(y)\simeq \pi + Cy$ for $|y|\ll1$, where $C$ is a constant
(Fig. \ref{fi:g} shows how well this approximation holds). Thus the
mapping (\ref{eq:mapone})--(\ref{eq:mapthree}) reduces to 
\begin{eqnarray} 
y_n' &= & y_n-\half\epsilon\sin m\phi_n,\nonumber \\
\phi_{n+1} &= & \phi_n + \pi + Cy_n',\nonumber \\
y_{n+1} &= & y_n'-\half\epsilon\sin m\phi_{n+1}.
\end{eqnarray}
If in addition $m$ is even, we can change variables to $q = m\phi + \pi $, 
$p = mCy$, to get
\begin{eqnarray}
p_n' &= & p_n+\half K\sin q_n,\nonumber\\
q_{n+1}&= &q_n+ p_n',\nonumber \\
p_{n+1}&= &p_n'+\half K\sin q_{n+1},
\end{eqnarray}
where $K = mC\epsilon$. In this form, the map is the symmetric expression of
the Chirikov-Taylor map, otherwise known as the standard map. This map serves
dynamicists as a laboratory for examining the behavior of area-preserving maps
as one increases the strength of perturbations. We refer the reader to the
vast literature on the standard map (e.g. Lichtenberg \& Lieberman 1992), and
simply recall the transition to global stochasticity that occurs as $K$
exceeds the critical value of $\simeq 1$, or equivalently when 
\be 
\epsilon\simeq \frac{1}{mC}.
\label{eq:stand}
\ee 
Thus we expect that the map is regular when the
non-axisymmetric perturbation is small, but only in the range
$1<\alpha<2$. More specifically, we shall find below that when $\alpha<1$
near-radial orbits can exhibit a complex chaotic structure, even for
arbitrarily small non-axisymmetric perturbations. 

\subsection{Non-axisymmetric potentials}\label{sec:nonaxipot}

We now ask what scale-free non-axisymmetric potentials can be
generated by plausible density distributions. Scale-free density
distributions can be written in the form
\be
\rho({\bf r})=r^{\alpha-2}\sum_{\ell,m}a_{\ell m}\ylm.
\ee
The corresponding potential is (e.g. Binney \& Tremaine 1987, \S2.4)
\be
\Phi({\bf r})=-4\pi Gr^\alpha\sum_{\ell,m}{a_{\ell m}\ylm\over
(\ell-\alpha)(\alpha+\ell+1)},\qquad -1-\ell<\alpha<\ell.
\label{eq:multi}
\ee 
When $\ell=0$, the range of validity of equation (\ref{eq:multi}) can be
extended to all $\alpha>0$, since the dynamics are determined by the
radial force $-d\Phi/dr$, which can be determined from (\ref{eq:multi})
whenever $-1<\alpha$.  Outside the range in which equation (\ref{eq:multi}) is
valid, the multipole potential is determined by the distribution of material
at very large or very small radii, so that it satisfies Laplace's equation and
$\alpha=\ell$ or $-\ell-1$. Thus the range of potentials in which we are
interested is given by 
\begin{eqnarray}
-1\le\alpha, &\quad& \ell=0,\nonumber\\
-1-\ell\le\alpha\le\ell,&\quad & \ell>0.
\label{eq:kkkk}
\end{eqnarray}
Since we restrict ourselves to the range of exponents $-1\le\alpha\le2$, these
constraints are always satisfied for monopole, quadrupole and higher multipoles
($\ell=0$ or $\ell\ge2$). For dipole potentials ($\ell=1$), the constraints
are only satisfied in the smaller range $-2\le\alpha\le1$.

To assess the realism of the map, we shall compare trajectories in the map to
orbits in scale-free non-axisymmetric potentials of the form 
\be 
\Phi_\alpha(x_1,x_2)=\left\{\begin{array}{cc}\hbox{sgn}(\alpha)
\left(x_1^2+{\textstyle x_2^2\over \textstyle
b^2}\right)^{\half\alpha}, &\alpha\not=0, \\ \half
\log\left(x_1^2+{\textstyle x_2^2\over
\textstyle b^2}\right), &\alpha=0, \end{array} \right.
\label{eq:potdefna}
\ee 
where $x_1=r\cos\phi$ and $x_2=r\sin\phi$ are the usual Cartesian
coordinates. The potential is specified by the exponent $\alpha$ and the axis
ratio of the equipotentials, $b\le 1$. The maximum angular momentum for an
orbit of given energy now depends on the azimuthal angle $\phi$, and is given
in terms of the maximum angular momentum in the analogous axisymmetric
potential (eq. \ref{eq:ellmax}) by
\be L_{\rm
max}(E,\phi)={L_c(E)\over\big(\cos^2\phi+\sin^2\phi/b^2\big)^{1/2}};
\quad\hbox{thus}\quad |y|=\left|L\over L_c\right|\le
\big(\cos^2\phi+\sin^2\phi/b^2\big)^{-1/2}.
\label{eq:fdef}
\ee

The map that approximates motion in this potential is specified by the same
$\alpha$, azimuthal wavenumber $m=2$, and the torque amplitude $\epsilon$.  To
relate $\epsilon$ to the axis ratio $b\equiv 1-\delta$, we examine near-radial
orbits ($y\ll1$) in nearly axisymmetric potentials ($\delta\ll1$). Then it is
straightforward to show that to lowest order in $\delta$ and $y$
\be
{\epsilon\over\delta}=Q_\alpha\equiv \left\{\begin{array}{cc}
\left(\textstyle 2\pi\over\textstyle\alpha\right)^{1/2}
\left(1+\half\alpha\right)^ {{\textstyle
{1\over\alpha}}+\half}{\textstyle\Gamma(1+1/\alpha) \over
\textstyle \Gamma({3\over 2}+1/\alpha)},
&\alpha>0, \\ (2\pi e)^{1/2},&\alpha=0, \\
\left(\textstyle 2\pi\over\textstyle\beta\right)^{1/2}
\left(1-\half\beta\right)^ {{\textstyle
-{1\over\beta}}+\half}{\textstyle\Gamma(-\half+1/\beta) \over
\textstyle\Gamma(1/\beta)},
&\alpha=-\beta<0. \\
\end{array} \right.
\label{eq:delepsrat}
\ee
This ratio varies smoothly from $Q_2=\pi=3.1416$, to $Q_0=4.1327$, to 
$Q_{-1}=2\pi=6.2832$.

The map is based on the approximation that torques are concentrated near
apocenter---that the intervals of the orbit when torques are
exerted (mostly near apocenter) and when the orbit precesses (mostly near
pericenter) are disjoint. This approximation is most accurate for nearly
radial orbits---if the orbit is almost exactly radial then all of the
precession occurs as the orbit passes the origin. If the non-axisymmetric
component of the potential is given by (\ref{eq:potdefna}), then the map
is most accurate for concave potentials ($\alpha>0$), since in this
case the torque is concentrated at large radii.

A second possible form for the potential is
\be
\Phi_\alpha(x_1,x_2)=c(x_2^2-x_1^2)+\left\{\begin{array}{cc}\hbox{sgn}(\alpha)
\left(x_1^2+x_2^2\right)^{\half\alpha}, &\alpha\not=0, \\ \half
\log\left(x_1^2+x_2^2\right), &\alpha=0; \end{array} \right.
\label{eq:potdefnab}
\ee 
this form is not scale-free, but may be more natural if the
non-axisymmetric potential arises from external tidal forces. For this
potential the torques are always concentrated at large radii.

\subsection{The logarithmic potential} \label{sec:log}

We first apply the map to the logarithmic potential ($\alpha=0$), which is
relevant to realistic galaxy models, and also represents a boundary 
between potentials in which near-radial orbits precess by
$\pi$ in one orbit and those that precess by smaller angles
(eq. \ref{eq:gzero}). The behaviour of orbits in this potential has been
examined by Richstone (1980, 1982), \cite{binspe82}, \cite{gerbin85},
\cite{pfedez88}, and \cite{mirsch89}. 

\begin{figure}
\centerline{\hbox{\epsfxsize=3.5in\epsfbox{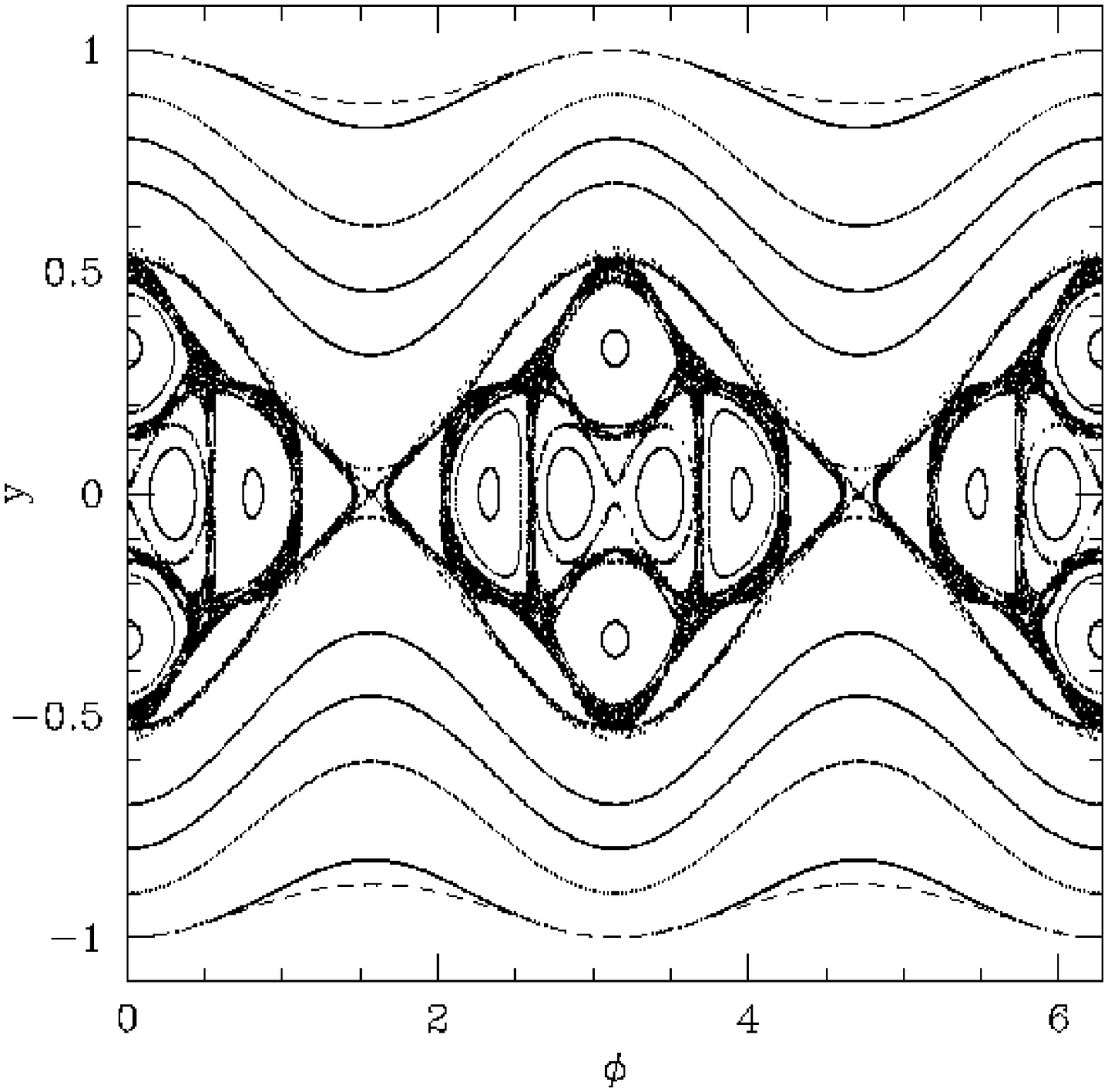}}}
\centerline{\hbox{\epsfxsize=3.5in\epsfbox{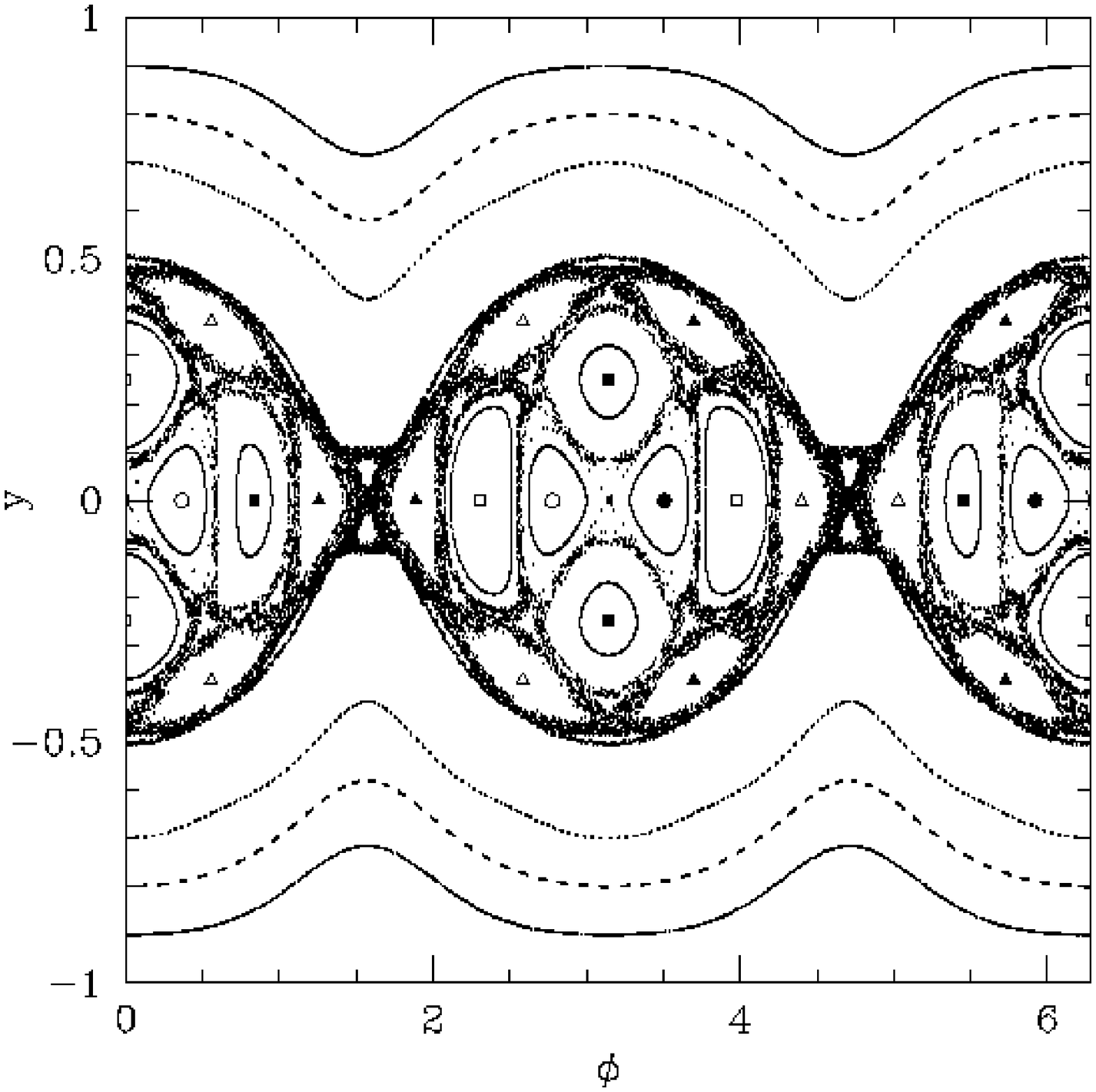}}}
\caption[Figure 2]{(a) Surface of section (SOS) for orbits in the potential
(\ref{eq:potdefna}), with axis ratio $b=0.88$. Each orbit is plotted for 1000
iterations, except for one stochastic orbit, which is plotted for 30,000
iterations to improve the definition of the stochastic web.  The vertical axis
is the dimensionless angular momentum (\ref{eq:ydef}) and the horizontal angle
is the apocenter azimuth. The dashed lines show the maximum dimensionless
angular momentum allowed by (\ref{eq:fdef}). (b) The map 
(\ref{eq:mapone}--\ref{eq:mapthree}) with parameters $\alpha=0$, $m=2$,
$\epsilon=0.5$. Resonant orbits of period 2 (``bananas''), 4 (``fish''), and 6
(``pretzels'') are marked by circles, squares, and triangles on the map; 
open and filled symbols denote different orbits.}
\label{fi:isomap}
\end{figure}

Figure \ref{fi:isomap}a shows the surface of section (SOS) for orbits in the
potential (\ref{eq:potdefna}) with $\alpha=0$ and $b=0.88$. The SOS shows the
dimensionless angular momentum and azimuth each time the orbit passes through
apocenter. The dashed lines show the maximum dimensionless angular momentum
allowed by (\ref{eq:fdef}). The analogous map, with azimuthal wavenumber $m=2$
and dimensionless strength $\epsilon=Q_0(1-b)=0.5$, is shown in Figure
\ref{fi:isomap}b. The principal differences are: (i) the shapes of the regular
orbits and the chaotic zones are slightly different; (ii) small resonant
islands for near-circular orbits are present in the integration but not the
map (between the last plotted orbit and the dashed boundary, centered on
$\phi=\half\pi,\frac{3}{2}\pi$); (iii) in the map there is no maximum angular
momentum, so that some orbits continue to values of $|y|>1$ (these are not
shown since they have no physical relevance). These differences are fairly
minor, and mostly occur for near-circular orbits, for which the map was never
intended to work well. Overall, the remarkable similarity of the two figures
shows that the map correctly captures the qualitative dynamics of eccentric
orbits in the logarithmic potential.

The SOS contains three main types of orbit:

\begin{itemize}

\item Loop orbits: these cross $\phi=0$ with angular momentum
$|y|>0.5$. For each orbit sgn$(y)$ is fixed; that is, they circulate around
the center in a fixed sense. The boundary of the loop orbits is defined by a
separatrix that passes through the unstable fixed points at
$\phi={1\over2}\pi,{3\over2}\pi$, $y=0$.  

\item Box-like orbits: Orbits with zero average angular momentum, which occupy
a set of resonant islands inside the separatrix. The orbits are box-like
because the apocenter azimuth librates around $\phi=0$ or $\pi$ and avoids
$\phi=\half\pi$ and $\frac{3}{2}\pi$. These orbits were examined by
\cite{gerbin85}, \cite{pfedez88}, and christened ``boxlets'' by
\cite{mirsch89}. Each periodic boxlet has a signature which is the sequence of
the signs of the angular momentum at successive apocenters ($+,-$ or 0). The
periodic orbits at the centers of the major islands correspond to ``banana''
orbits (period 2, signature $[00]$), ``fish'' orbits (period 4, signature
$[0+0-]$), and ``pretzel'' orbits (period 6, signature
$[0++\,0--]$); the nomenclature is that of \cite{bin82} and \cite{mirsch89}. 

\item A stochastic web, which shows up on the SOS as a connected fuzzy
structure passing through $y=0$, $\phi=\half\pi,\frac{3}{2}\pi$. Despite the
complicated geometry of the web, this is a single orbit, which was followed
for 30,000 iterations (compared to 1000 for the other orbits) to improve the
coverage.

\end{itemize}

The SOS also provides preliminary information on which orbits are needed to
construct self-consistent equilibrium galaxy models. The torques described by
equation (\ref{eq:mapone}) with $m=2$, or the potential (\ref{eq:potdefna}),
are generated by a density distribution that is elongated along the axis
$\phi=0,\pi$. The SOS shows that loop orbits are most eccentric when their
apocenters are at $\phi=\half\pi,\frac{3}{2}\pi$ and hence these do not
support the required elongation of the density distribution. The box-like
orbits, however, have apocenters that librate around $\phi=0$ or $\pi$ and
avoid $\half\pi$ and $\frac{3}{2}\pi$; therefore these are aligned with the
required figure. However, detailed studies of orbits in the logarithmic
potential are required to determine whether the density distribution is narrow
enough to allow self-consistent equilibrium models (Richstone 1980, Pfenniger
\& de Zeeuw 1989, Lees \& Schwarzschild 1992, Kuijken 1993, Schwarzschild
1993).

Let us now consider the stochastic web in more detail. Because of the
singularity in the logarithmic potential at $r=0$, we expect that any orbit
passing close to the origin will be chaotic. In the context of the map, an
orbit leaving the origin has zero angular momentum and azimuthal angle $\phi$;
shortly before apocenter it receives an angular momentum kick
$-\half\epsilon\sin m\phi$ (cf. eq. \ref{eq:mapone}); therefore its
phase-space coordinates at apocenter are 
\be
(\phi,y)=(\phi_0,-\half\epsilon\sin m\phi_0).
\label{eq:webst}
\ee
All points in the map satisfying this constraint---and their iterates
under the map---should belong to stochastic orbits.  What is less obvious is
that they all belong to the {\it same} orbit, i.e. that all orbits passing
through the origin are part of the same stochastic web. We cannot prove this
result but it is consistent with all of our numerical experiments with
logarithmic potentials.

\begin{figure}
\centerline{\hbox{\epsfxsize=3.5in\epsfbox{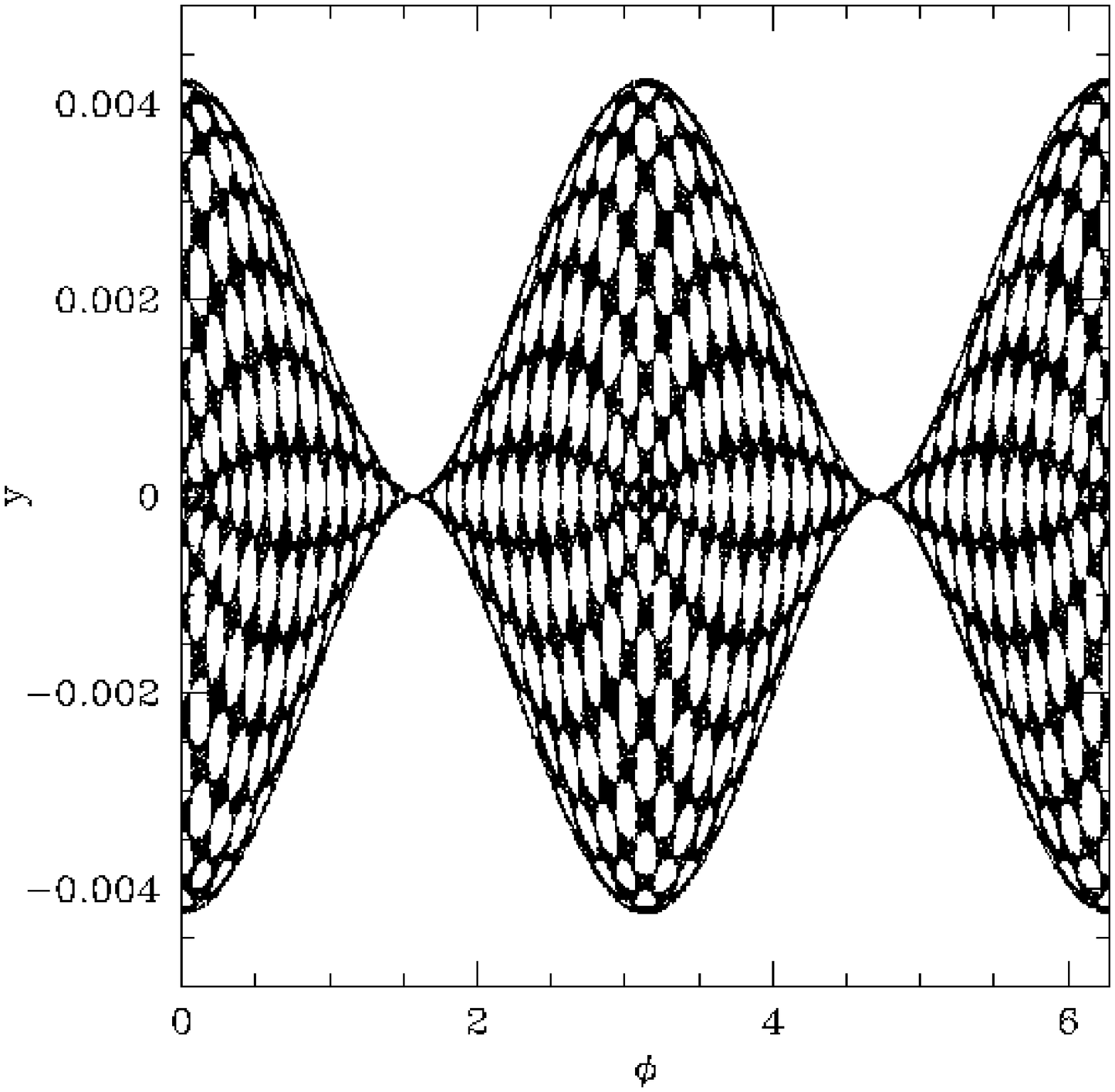}}}
\centerline{\hbox{\epsfxsize=3.5in\epsfbox{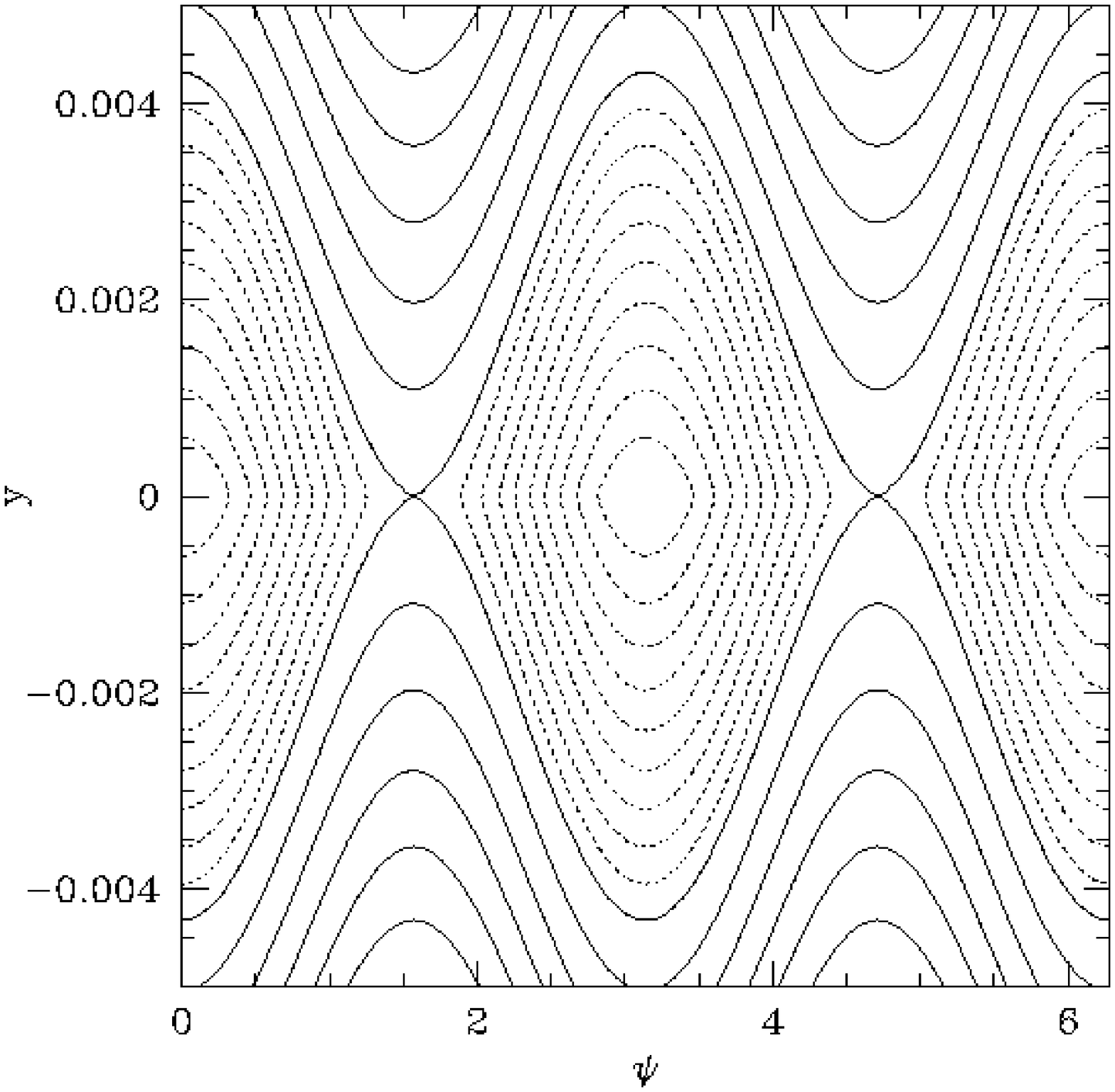}}} 
\caption[Figure 3]{(a) A single
stochastic orbit in the map (\ref{eq:mapone}--\ref{eq:mapthree}) with
parameters $\alpha=0$, $m=2$, $\epsilon=0.001$. The orbit has initial
conditions $\phi={1\over4}\pi$, $y=-0.0005$ (cf. eq. \ref{eq:webst}) and is
iterated 100,000 times. (b) Level surfaces of the averaged Hamiltonian
(\ref{eq:hamdefav}), for the same parameters as in (a). The variable $\psi$
equals $\phi$ or $\phi\pm\pi$. Level surfaces corresponding to loop orbits are
shown by solid lines, and those corresponding to box-like orbits are shown by
dotted lines.}
\label{fi:isoweb}
\end{figure}

The geometry of the web is particularly striking when $\epsilon$ is
small. Figure \ref{fi:isoweb}a shows the web generated by 100,000 iterations of
a single orbit in a map with $\alpha=0$, $m=2$, $\epsilon=0.001$. A very
similar plot is generated by direct orbit integration in the analogous
non-axisymmetric potential.

The following analysis provides some insight into the behavior shown in Figure
\ref{fi:isoweb}a. Let $\psi_n=\phi_n-ng_0(\alpha)$ mod$(2\pi)$. Since
$g_0(\alpha)=\pi$ for $\alpha\ge0$, $\psi_n=\phi_n$ if $n$ is even and
$\psi_n=\phi_n\pm\pi$ if $n$ is odd. Then for $m=2$ and $\alpha\ge0$ equations
(\ref{eq:mapone}--\ref{eq:mapthree}) become
\begin{eqnarray}
y_n' & = & y_n-\half\epsilon\sin 2\psi_n,\nonumber \\
\psi_{n+1} & = & \psi_n+g(\alpha,y_n')-g_0(\alpha), \nonumber \\
y_{n+1} & = & y_n'-\half\epsilon\sin 2\psi_{n+1}. 
\label{eq:mapfive}
\end{eqnarray}
This mapping is equivalent to motion under the Hamiltonian
\be
H=G(\alpha,y)-\half\epsilon\delta_1(t)\cos 2\psi
=G(\alpha,y)-\half\epsilon\cos 2\psi\sum_{j=-\infty}^\infty\exp(2\pi i j t).
\label{eq:hamdef}
\ee
Here $\delta_1(t)=\sum_{k=-\infty}^\infty\delta(t-k)$ is the periodic
delta-function with unit period, $t$ is a continuous variable which equals $n$
at the $n^{\rm th}$ iteration of the mapping, $y$ is the momentum conjugate to
the coordinate $\psi$, $\psi(t=n)=\psi_n$, $y(t)=y_n'$ for $n<t<n+1$, and
\be
G(\alpha,y)=\int_0^y[g(\alpha,y)-g_0(\alpha)]dy.
\ee
If the motion is slow, the terms in (\ref{eq:hamdef}) with $j\not=0$ have
relatively little effect, and we can approximate the Hamiltonian by its
averaged value 
\be
\overline H=G(\alpha,y)-\half\epsilon\cos 2\psi.
\label{eq:hamdefav}
\ee 
The motion is along level surfaces of this Hamiltonian, which are plotted
in Figure \ref{fi:isoweb}b for the same values of $\alpha$ and $\epsilon$ as
the map in Figure \ref{fi:isoweb}a. We see that the averaged Hamiltonian
correctly captures the location of the separatrix and the division into loop
orbits and box-like orbits (but not, of course, the detailed structure of the
stochastic web). 

\subsection{The Liapunov exponent}

The map can be used to determine the Liapunov exponent associated with the web
for any value of the perturbation strength $\epsilon$. We simply follow the
tangent map for an orbit that passes close to the center, i.e. one with
initial coordinates given by equation (\ref{eq:webst}). 
The tangent map is defined by
taking differentials of equations (\ref{eq:mapone}--\ref{eq:mapthree}):
\begin{eqnarray}
dy_n' & = & dy_n-\half m\epsilon\cos(m\phi_n)d\phi_n,\nonumber \\
d\phi_{n+1} & = & d\phi_n+{\partial g\over\partial y}(\alpha,y_n')dy_n',
\nonumber \\
dy_{n+1} & = & dy_n'-\half m\epsilon\cos m\phi_{n+1}d\phi_{n+1}. 
\end{eqnarray}
The Liapunov exponent is
\be
\lambda=\lim_{n\to\infty}{1\over n}\ln
   \left(d\phi_n^2+dy_n^2\over d\phi_0^2+dy_0^2\right)^{1/2}.
\label{eq:liapdef}
\ee
In practice we estimate the Liapunov exponent using large but finite
$n$. Since we expect that all initial conditions (\ref{eq:webst}) lie in the
web (i.e. they are all the same orbit), the Liapunov exponent should be
independent of the initial azimuth $\phi_0$.

\begin{figure}
\centerline{\hbox{\epsfxsize=4in\epsfbox{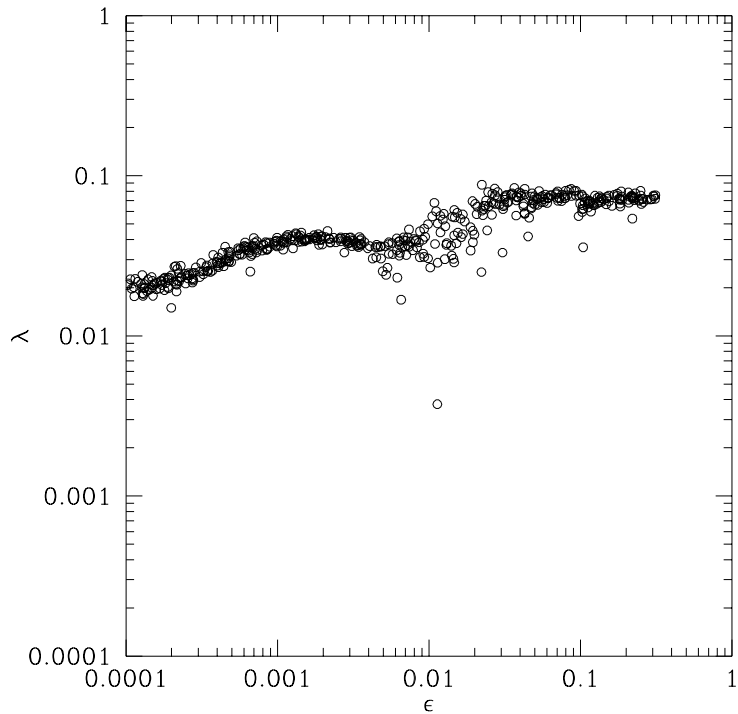}}}
\caption[Figure 4]{The Liapunov exponent $\lambda$ (eq. \ref{eq:liapdef}) as a
function of perturbation strength $\epsilon$, for maps with $\alpha=0$ and
$m=2$. The initial conditions were taken from (\ref{eq:webst}) with $\phi_0$
chosen randomly in $[0,2\pi)$. The perturbation strength $\epsilon$ was also
chosen at random, from a uniform distribution in $\log\epsilon$ between $-4$
and $-0.5$. A total of 500 ($\phi_0,\epsilon)$ pairs were examined, and each
was iterated $10^5$ times.}
\label{fi:liap}
\end{figure}

Figure \ref{fi:liap} shows the Liapunov exponent $\lambda$ for maps with
$\alpha=0$ and $m=2$. The initial azimuth $\phi_0$ was chosen randomly in the
interval $[0,2\pi)$ and each map was iterated 100,000 times. Although there
are some outliers, in general the scatter is small, confirming that $\lambda$
is almost independent of $\phi_0$. What is unexpected is that $\lambda$
depends only weakly on the perturbation strength $\epsilon$---from around 0.07
near $\epsilon=0.3$ to $0.02$ near $\epsilon=10^{-4}$. 

\begin{figure}
\centerline{\hbox{\epsfxsize=3in\epsfbox{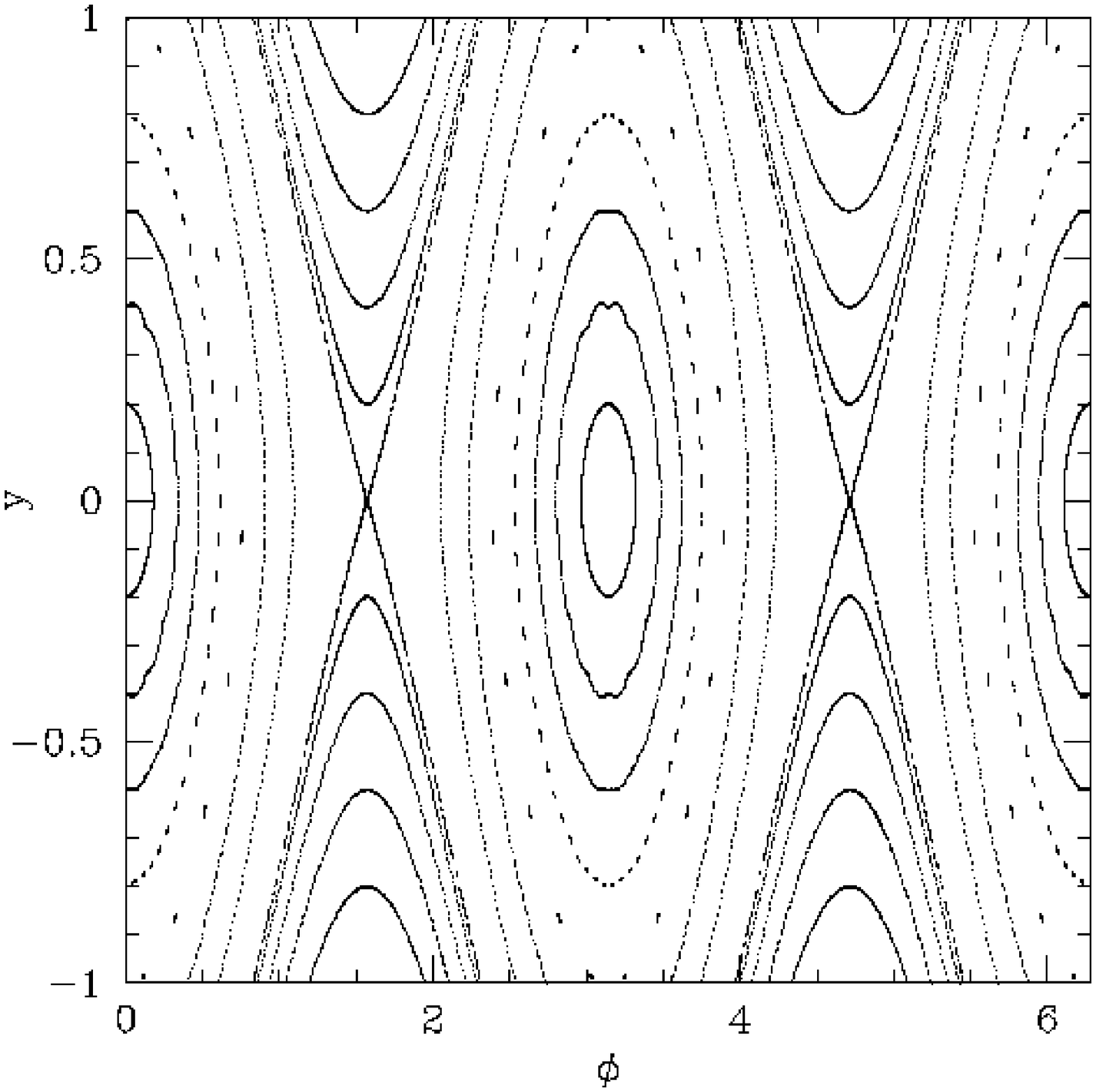}
                  \epsfxsize=3in\epsfbox{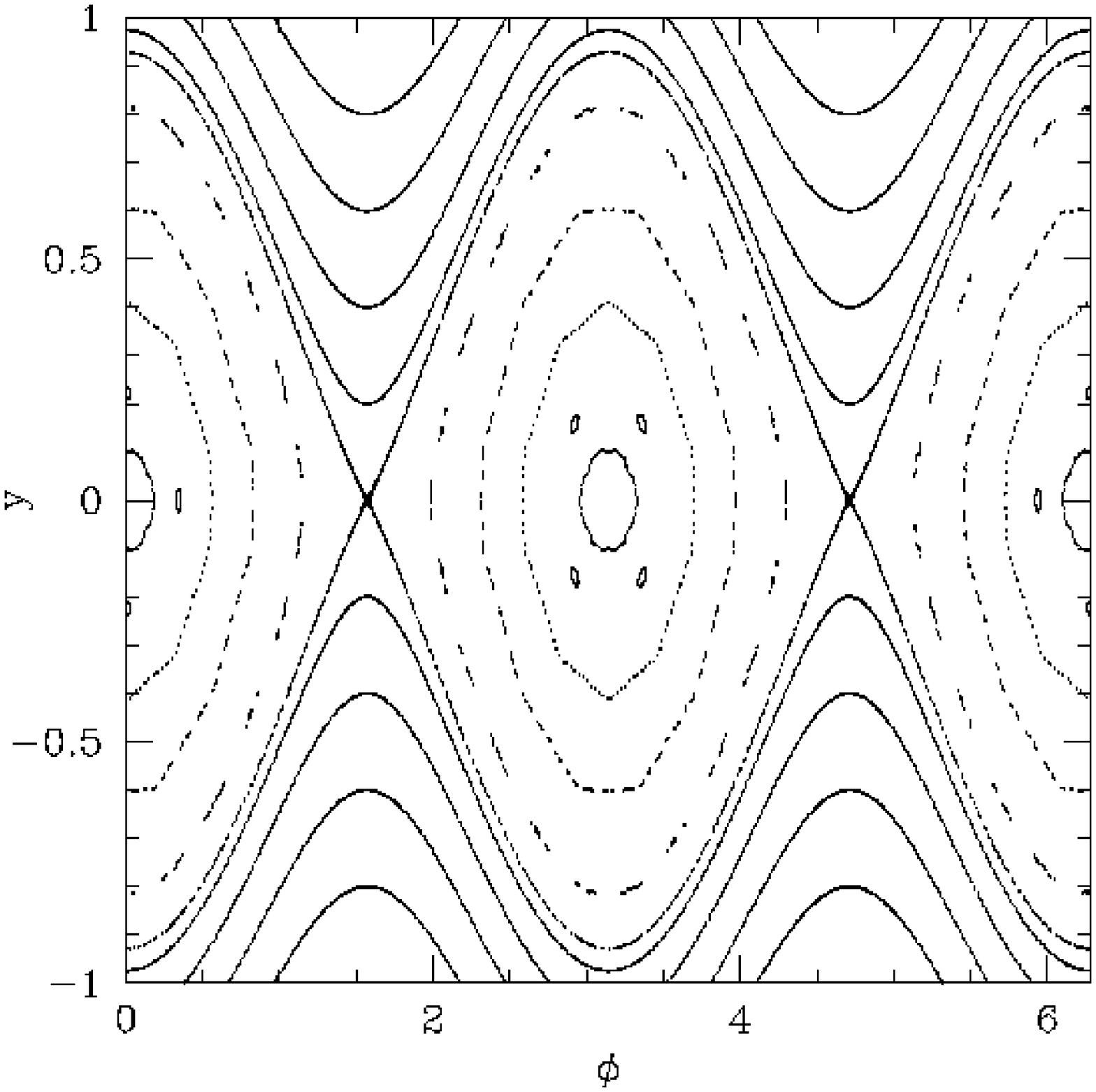}}}
\centerline{\hspace*{1.45in}$(a)$\hfill$(b)$\hspace{1.3in}}
\centerline{\hbox{\epsfxsize=3in\epsfbox{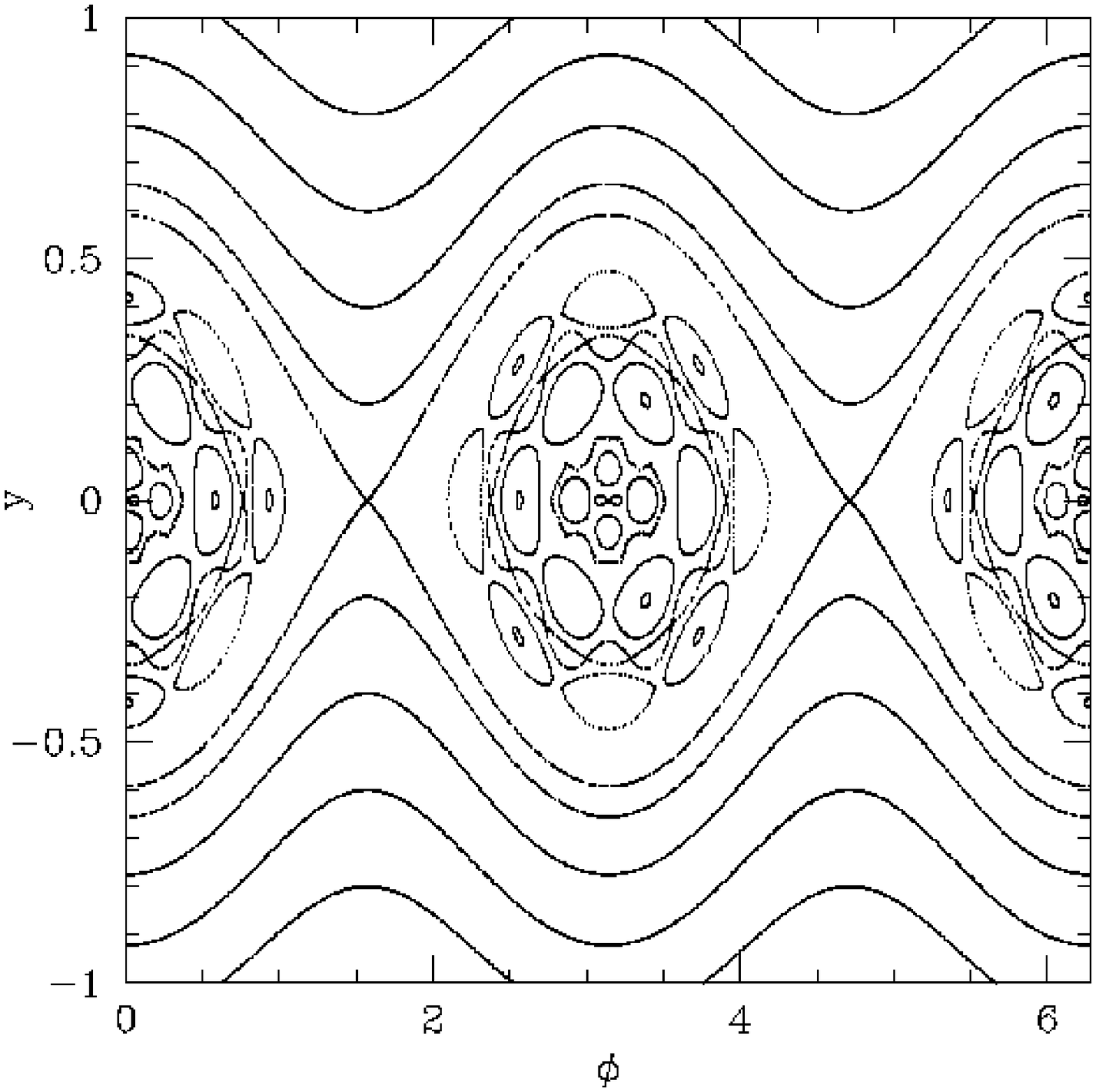}
                  \epsfxsize=3in\epsfbox{ 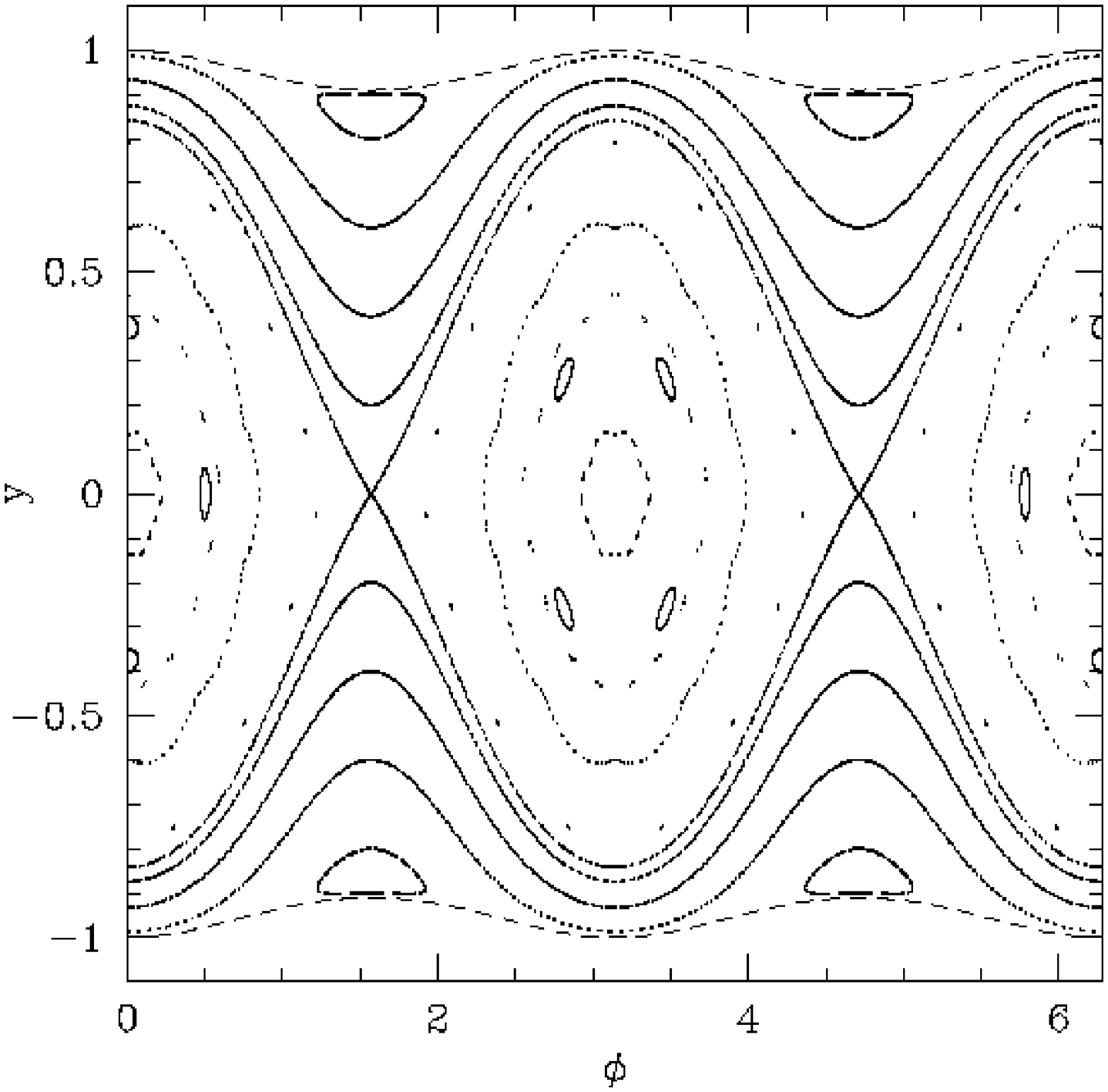}}}
\centerline{\hspace*{1.45in}$(c)$\hfill$(d)$\hspace{1.3in}}
\caption[Figure 5]{The map
(\ref{eq:mapone}--\ref{eq:mapthree}) with parameters $m=2$, $\epsilon=0.3$,
and (a) $\alpha=1.5$; (b) $\alpha=1$; (c) $\alpha=0.5$. The orbits have been
iterated 1000 times, except for the orbits at the separatrices, which are
iterated 5000 times to provide better coverage.  (d) The SOS for orbits in the
potential (\ref{eq:potdefna}), with $\alpha=1$ and axis ratio $b=0.91$
(cf. eq. \ref{eq:delepsrat}). The dashed lines show the maximum dimensionless
angular momentum allowed by (\ref{eq:fdef}). Each orbit is plotted for 1000
periods, except for the separatrix orbit, which is plotted for 5000 periods. }
\label{fi:conc}
\end{figure}

\section{Concave potentials ($\alpha>0$)}

The maps with $m=2$, $\epsilon=0.3$ and $\alpha=1.5,1,0.5$ are shown in Figure
\ref{fi:conc}a,b,c. Like the map for the logarithmic potential examined
already, these exhibit three types of orbits: loop orbits, which have a fixed
sign of angular momentum; box-like orbits, which have zero average angular
momentum and an apocenter azimuth that librates about $\phi=0$ or $\pi$; and a
separatrix orbit, which passes through $y=0$, $\phi=\half\pi,\frac{3}{2}\pi$
and separates the loop orbits from the box-like orbits.

Panel (d) shows the SOS for orbits integrated in the potential
(\ref{eq:potdefna}) with $\alpha=1$, $b=0.91$, which corresponds to the map
(b). The similarity of the map in panel (b) to the SOS in panel (d) confirms
that the map captures the qualitative dynamics in the non-axisymmetric
potential. The principal difference, apart from minor changes in orbit shape,
is the small resonant islands for near-circular orbits that appear in the
integration but not the map (near the dashed boundary, centered on
$\phi=\half\pi,\frac{3}{2}\pi$).

\begin{figure}
\centerline{\hbox{\epsfxsize=4in\epsfbox{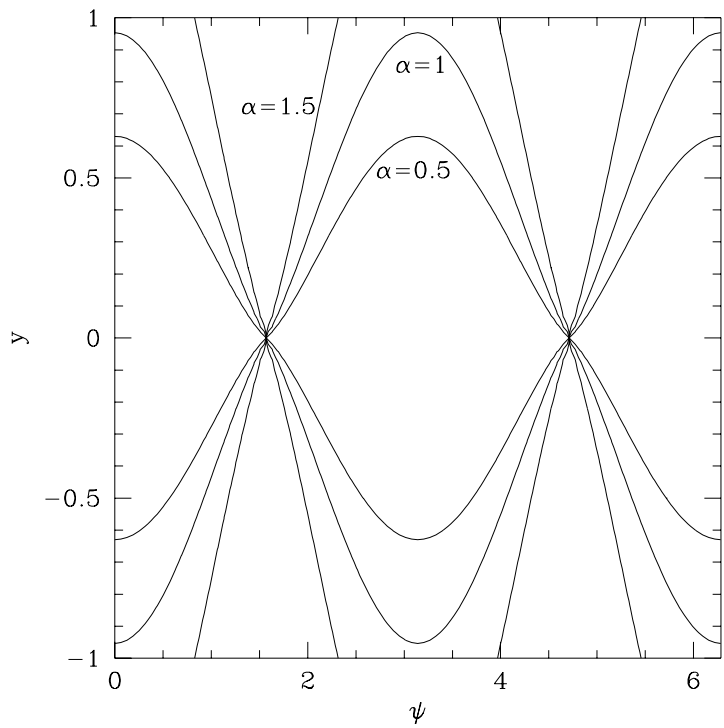}}}
\caption[Figure 6]{Separatrices of the averaged Hamiltonian
(\ref{eq:hamdefav}), for $m=2$, $\epsilon=0.3$, and $\alpha=1.5,1,0.5$. These
agree well with the locations of the separatrix orbits in the corresponding
maps, panels (a),(b),(c) of Figure \ref{fi:conc}.  }
\label{fi:hamavv}
\end{figure}

The locations of the separatrices are well-described by the averaged
Hamiltonian (\ref{eq:hamdefav}), as illustrated in Figure \ref{fi:hamavv}.

There are two interesting transitions in the behaviour of the box-like
orbits as $\alpha$ is varied: (i) In Figures \ref{fi:conc}a,b
($\alpha\ge 1$) most of the box-like orbits form smooth curves
surrounding the points $y=0$, $\phi=0,\pi$, while in Figures
\ref{fi:conc}c and \ref{fi:isomap} ($0\le\alpha<1$) the box-like
orbits are almost all broken up into resonant islands. \cite{mirsch89}
call orbits of the first type ``normal boxes'' and those of the second
type ``boxlets''. Presumably this transition reflects the validity of
the standard map as an approximation to the present map when
$\alpha>1$ (\S \ref{sec:std}). (ii) In Figure \ref{fi:isomap} the
separatrix orbit is part of a stochastic web that surrounds all of the
resonant islands, while in Figure \ref{fi:conc}c the separatrix orbit
does not penetrate between the islands. Presumably this difference
arises because KAM surfaces divide the separatrix from the
islands in the latter case.

In \S\ref{sec:std} we argued that for $ 1 \leq
\alpha < 2$, the mapping resembles the Taylor-Chirikov map, with the critical
$\epsilon$ for global stochasticity given in equation (\ref{eq:stand}). Thus
the motion should be nearly regular for small $\epsilon$, in agreement with
the phase portraits shown in Figures \ref{fi:conc}a,b and d.

\subsection{A simpler map}

The behavior of box-like orbits in concave potentials can be clarified
by examining a simpler map. We begin with the map 
(\ref{eq:mapfive}) and assume that $\psi$ is small,
so that $\sin2\psi_n\simeq 2\psi_n$. We also replace
$g(\alpha,y)-g_0(\alpha)$ by its asymptotic form for small $y$
(eqs. \ref{eq:gasymp} and \ref{eq:ggasymp}). Thus we may write
\begin{eqnarray}
y_n'& = & y_n-\epsilon\psi_n,\nonumber\\
\psi_{n+1} & = & \psi_n+C\,\mbox{sgn}(y_n')|y_n'|^\beta,\nonumber \\
y_{n+1} &= & y_n'-\epsilon\psi_{n+1}.
\end{eqnarray}
where $C>0$ is a constant and $0<\beta<1$. 
We now re-scale the coordinates, choosing new variables 
$I\equiv (\epsilon C)^{1/(\beta-1)}y'$ and $\theta\equiv \epsilon
(\epsilon C)^{1/(\beta-1)}\psi$. In these coordinates the map takes the simpler
form 
\begin{eqnarray}
I_n' & = & I_n-\theta_n,\nonumber\\
\theta_{n+1} & = & \theta_n+ \mbox{sgn}(I_n')|I_n'|^\beta,\nonumber \\
I_{n+1} & = & I_n'-\theta_{n+1},
\label{eq:mapsimp}
\end{eqnarray}
The motion is governed by the Hamiltonian 
\be
H = \frac{|I|^{\gamma}}{\gamma} + {\delta}_1(t)
{\theta}^{2},
\ee
where ${\delta}_1(t)$ is the periodic Dirac delta function with unit period,
$\gamma=\beta+1$, $\theta_n$ is the value of $\theta(t)$ at $t=n$, and $I_n'$
is the value of $I(t)$ for $n<t<n+1$. We consider first the time-averaged
Hamiltonian
\be
H_{0} = |I|^{\gamma}/\gamma + {\theta}^{2}
\label{eq:H0}
\ee  
The corresponding phase-space portrait consists of closed curves that
are not differentiable on the axis $I=0$. With each trajectory we
associate the action:
\begin{eqnarray}
J &=& \frac{1}{2\pi} \oint I d\theta \nonumber \\
  &=& \frac{{\gamma}^{\frac{1}{\gamma}} W_\gamma}{\pi}
      H_0^{\frac{\gamma+2}{2\gamma}},
\end{eqnarray}
where $W_\gamma = {\int}_{-1}^{1} {(1 - x^{2})}^{\frac{1}{\gamma}}dx=
\pi^{1/2}\Gamma(1+1/\gamma)/\Gamma(\frac{3}{2}+1/\gamma)$. 
The energy is simply given by: $H_{0}(J) =
\omega_0J^{\frac{2\gamma}{\gamma+2}}$, with 
\be
\omega_0 = {1\over\gamma^{\frac{2}{\gamma+2}}}
\left(\frac{\pi}{W_\gamma}\right)^{\frac{2\gamma}{\gamma+2}}.
\ee
Note that the frequency $\omega(J) = dH_0/dJ$ diverges as $J$ tends to zero,
since $\gamma=\beta+1<2$. This feature implies that the motion is
highly unstable near the origin once we add the high-frequency terms of the
periodic delta function. 

The transformation to action-angle variables is 
accomplished with the generating function
\be
S(\theta,J)=\gamma^{\frac{1}{\gamma}}
\int^{\theta}{\big[H_{0}(J)-\theta^2\big]}^{\frac{1}{\gamma}} d\theta,
\ee
which naturally gives the angle conjugate to $J$, $\phi= \partial S/\partial
J$. 

Now we restrict ourselves to the case $\beta=\gamma-1 \ll 1$, which is valid
for potentials that are close to the logarithmic potential. We have
$\gamma\simeq1$, $H_0(J)=\omega_0J^{2/3}$, $\omega_0=(3\pi)^{2/3}2^{-4/3}$. We
can also write $\theta=(dH_0/dJ)^{-1}g(\phi)$, where $g(\phi)$ is the
$2\pi$-periodic sawtooth function defined by 
\be 
g(x)=\cases{x, & $|x|\le\half\pi$,\cr 
        \pi-x, & $\half\pi\le x\le\frac{3}{2}\pi$, \cr
g(2\pi+x).\cr} 
\ee 
In the action-angle variables $(J,\phi)$ of the averaged
Hamiltonian, the mapping Hamiltonian becomes 
\be 
H = H_0(J) +[\delta_1(t)-1]\left(dH_0\over dJ\right)^{-2}g^2(\phi) =
H_0(J)\left\{1+{4\over\pi^2}[\delta_1(t)-1]g^2(\phi)\right\}.  
\ee

We would like to analyze the principal resonances of this Hamiltonian.
After Fourier expansion of $g$ and $\delta$, the Hamiltonian takes the form

\be
H = H_0(J)\left[1 + \frac{2}{3}\sum_{n=1}^\infty \cos(2 \pi nt) + 
\frac{4}{{\pi}^{2}}\sum_{m\not=0}\sum_{n=1}^\infty
\frac{(-1)^{m}}{m^2}\cos2(m\phi-\pi nt)\right ] .
\ee
Choosing a resonant pair $(m, n)$, and including all terms of the form
$(pm,pn)$, we end up with the resonant Hamiltonian:
\be
H_{mn} = H_0(J)\left[1 + \frac{4}{{m^2\pi}^{2}}
\sum_{p=1}^\infty \frac{(-1)^{pm}}{p^2} \cos 2p(m\phi-\pi nt)\right].
\ee
We change coordinates with the canonical transformation: $F(J', \phi)
= 2J'(m\phi-\pi nt)$, giving $J = 2mJ'$ and ${\phi}' = 2(m\phi-\pi nt)$, and end
up with the one degree of freedom Hamiltonian 
\be
H'_{mn}(J', \phi')  =  H_0(2mJ') \left[1 + \frac{4}{{m^2\pi}^{2}} 
\sum_{p=1}^\infty \frac{(-1)^{pm}}{p^2} \cos(p\phi') \right] - 2\pi n J',
\ee
which simplifies to 
\begin{eqnarray}
H'_{mn}(J', \phi') & = & H_0(2mJ') \left[1 +
\frac{2}{3m^2}-\frac{2\phi'}{m^2\pi} +\frac{{\phi'}^2}{m^2\pi^2} \right]
- 2\pi n J',\quad \hbox{$0\le\phi'<2\pi$, $m$ even}\nonumber \\
& = & H_0(2mJ') \left[1 - \frac{1}{3m^2}+ \frac{{\phi'}^2}{m^2\pi^2} \right]
- 2\pi n J',\qquad\quad \hbox{  $-\pi\le\phi'\le\pi$, $m$ odd}.
\end{eqnarray}
This Hamiltonian has equilibria at $0$ and $\pi$. For $m$ even and $\gamma
\simeq 1$ we get   
\begin{eqnarray}
J_0     & = & \left[2 m \gamma\omega_0 \big(1+\frac{2}{3m^2}\big)\over 
\pi n(\gamma+2)\right]^{\frac{2+\gamma}{2-\gamma}},  \nonumber \\
J_{\pi} & = & \left[2 m\gamma\omega_0 \big(1-\frac{1}{3m^2}\big)\over
\pi n(\gamma+2)\right]^{\frac{2+\gamma}{2-\gamma}};
\label{eq:tori}
\end{eqnarray}
and for $m$ odd, $J_0$ and $J_{\pi}$ exchange values. The difference
between $J_0$ and $J_\pi$ diminishes with increasing $m$. The resonant
tori in $I$--$\theta$ space are recovered via equation (\ref{eq:H0}). 

\begin{figure}
\centerline{\hbox{\epsfxsize=6in\epsfbox{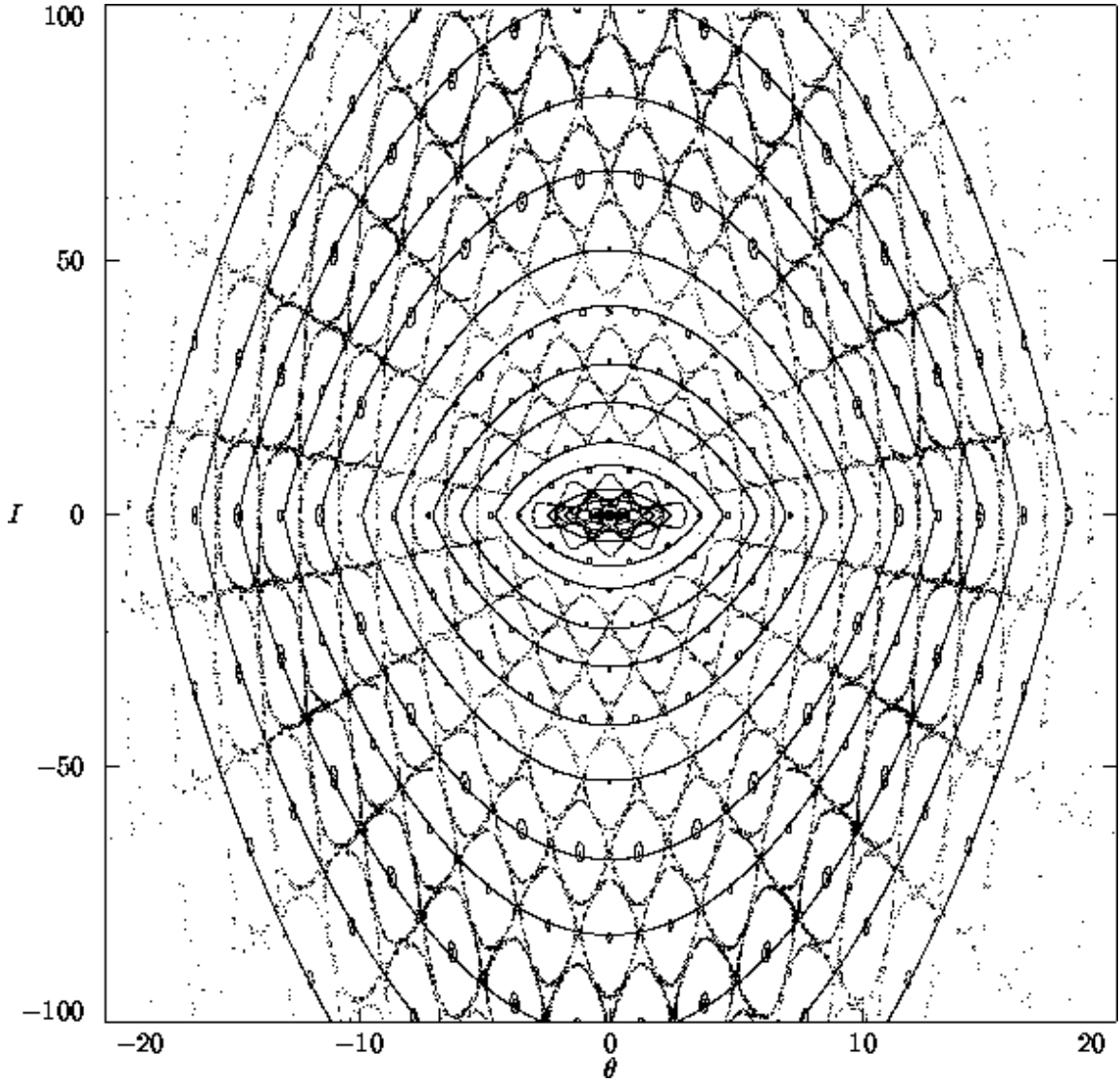}}}
\caption[Figure 7]{The resonant tori $J_\pi$ that are predicted by
(\ref{eq:tori}) for $n=1, m=1 \ldots 15$. The web determined by the map
(\ref{eq:mapsimp}) is superposed on top, along with one orbit near the center
of each of the major islands. The predicted tori pass through the periodic
points in the largest chains of islands outlined by the web.}
\label{fi:islands}
\end{figure}

We observe that, for $\beta \ll 1 $, the stochastic web outlines the
separatrices of the resonances corresponding to $(m > 0, n=1)$. In Figure
\ref{fi:islands}, we show, in solid lines, the resonant tori corresponding to
$J_\pi$ and $(m=1 \ldots 15, n=1)$, for $\beta=0.2$. Superposed on them is the
web structure that obtains at this value of $\beta$ from the map
(\ref{eq:mapsimp}). The tori string together the periodic points in chains of
islands. These periodic orbits and the ``boxlets'' associated with them
dominate the phase space. The separatrices, nested in tangent contact, form
the network along which orbits diffuse in the web. 

\section{Convex potentials ($\alpha<0$)}

\begin{figure}
\centerline{\hbox{\epsfxsize=3in\epsfbox{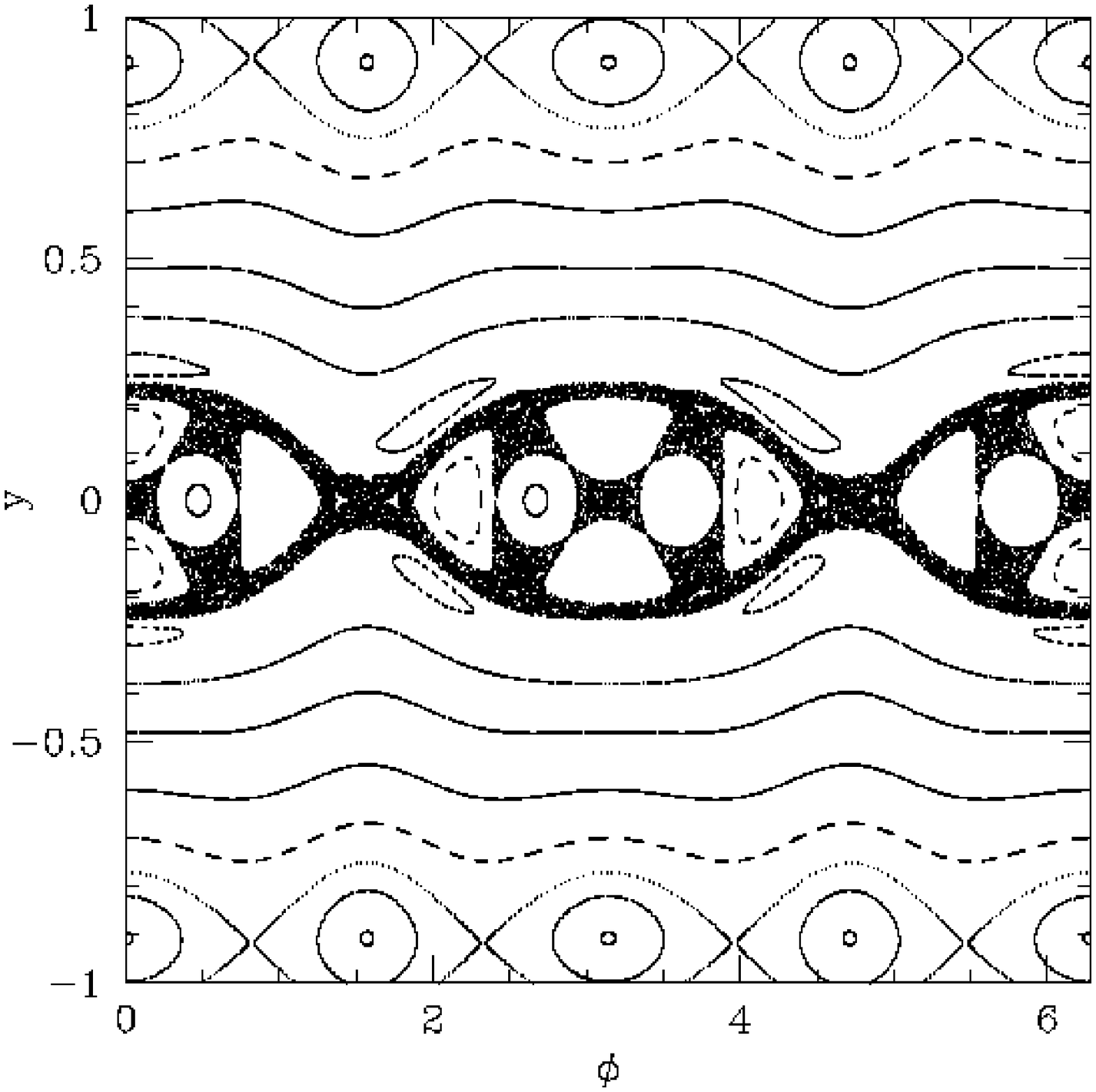}
                  \epsfxsize=3in\epsfbox{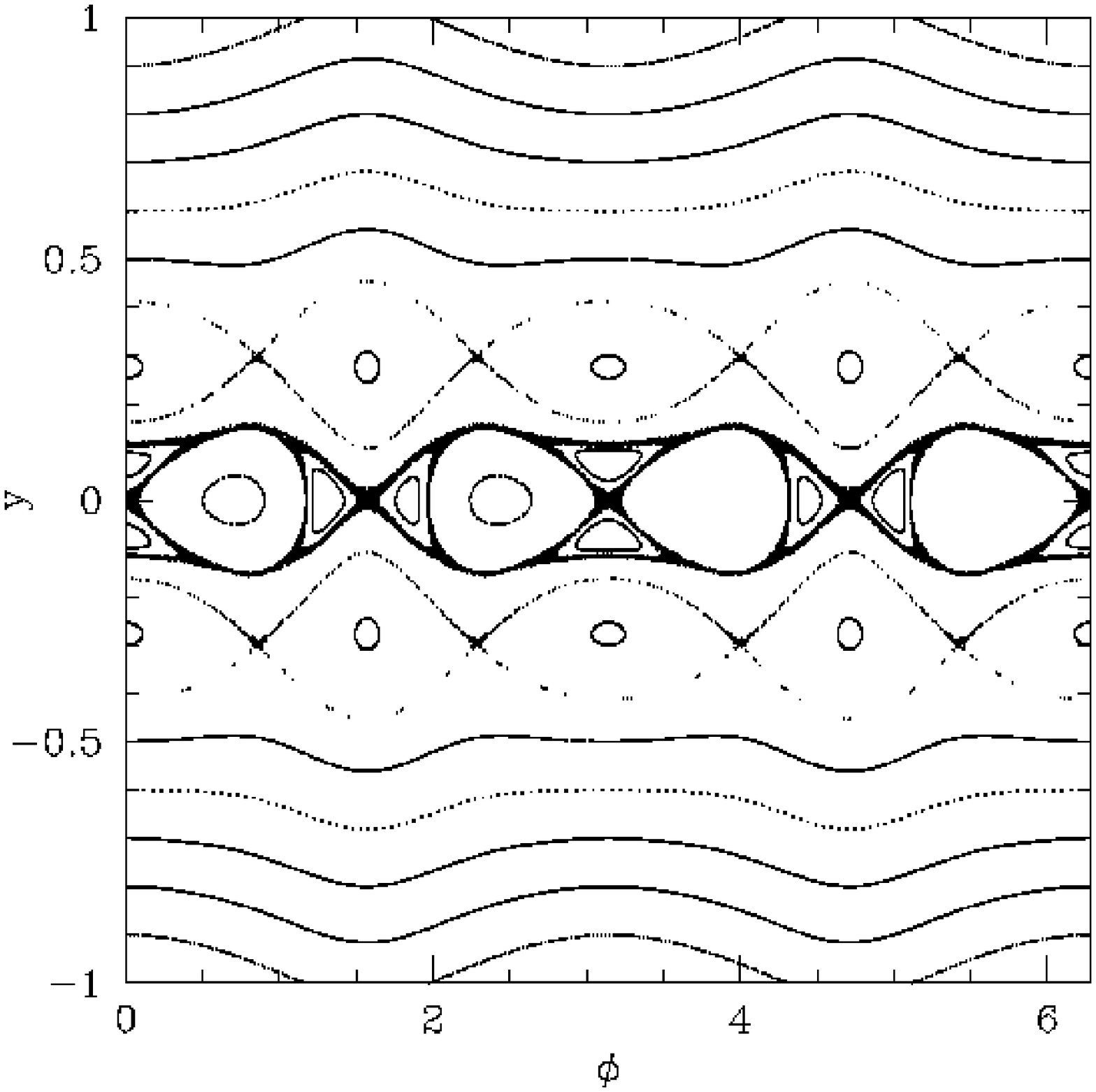}}}
\centerline{\hspace*{1.45in}$(a)$\hfill$(b)$\hspace{1.3in}}
\centerline{\hbox{\epsfxsize=3in\epsfbox{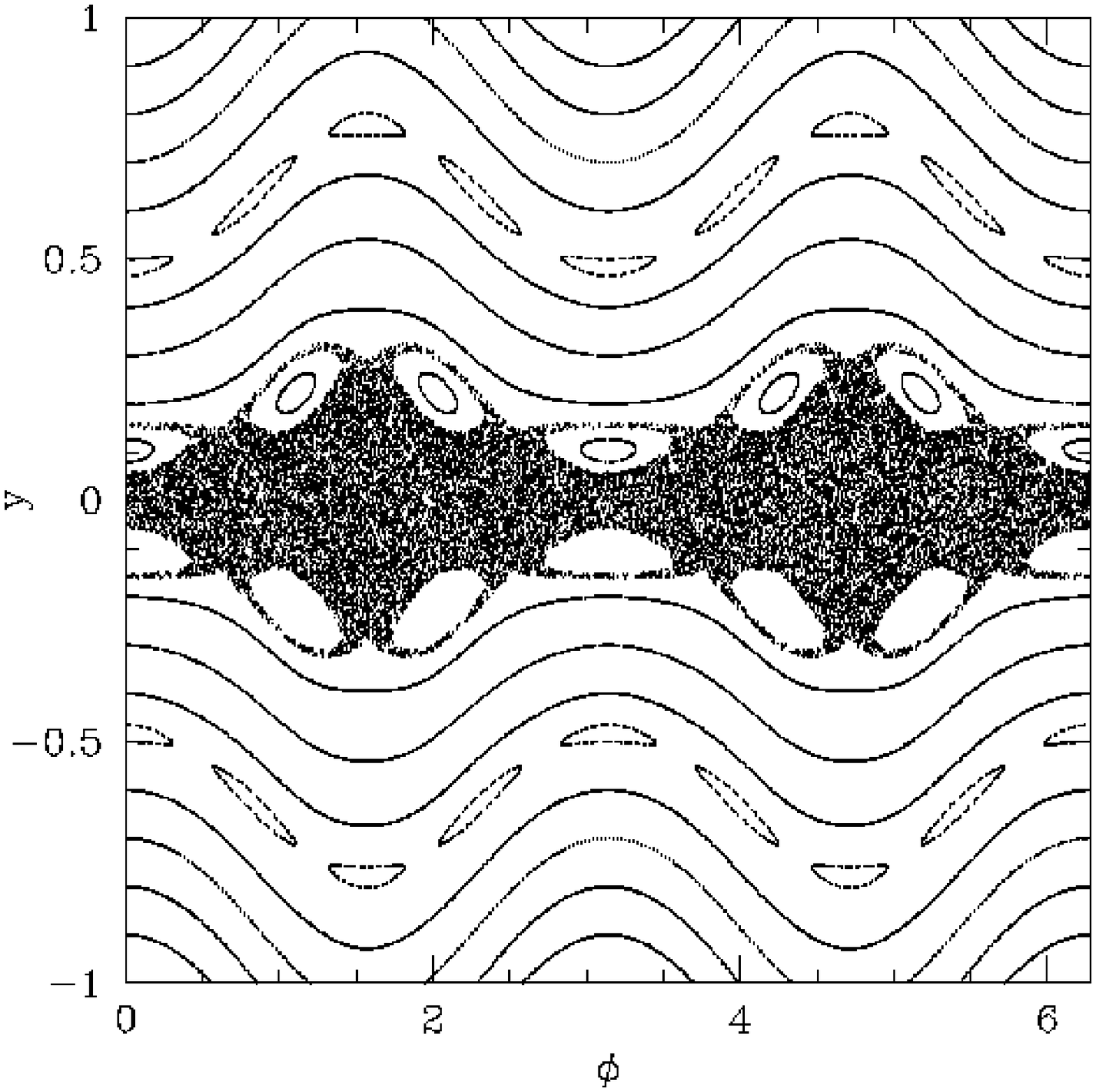}
                  \epsfxsize=3in\epsfbox{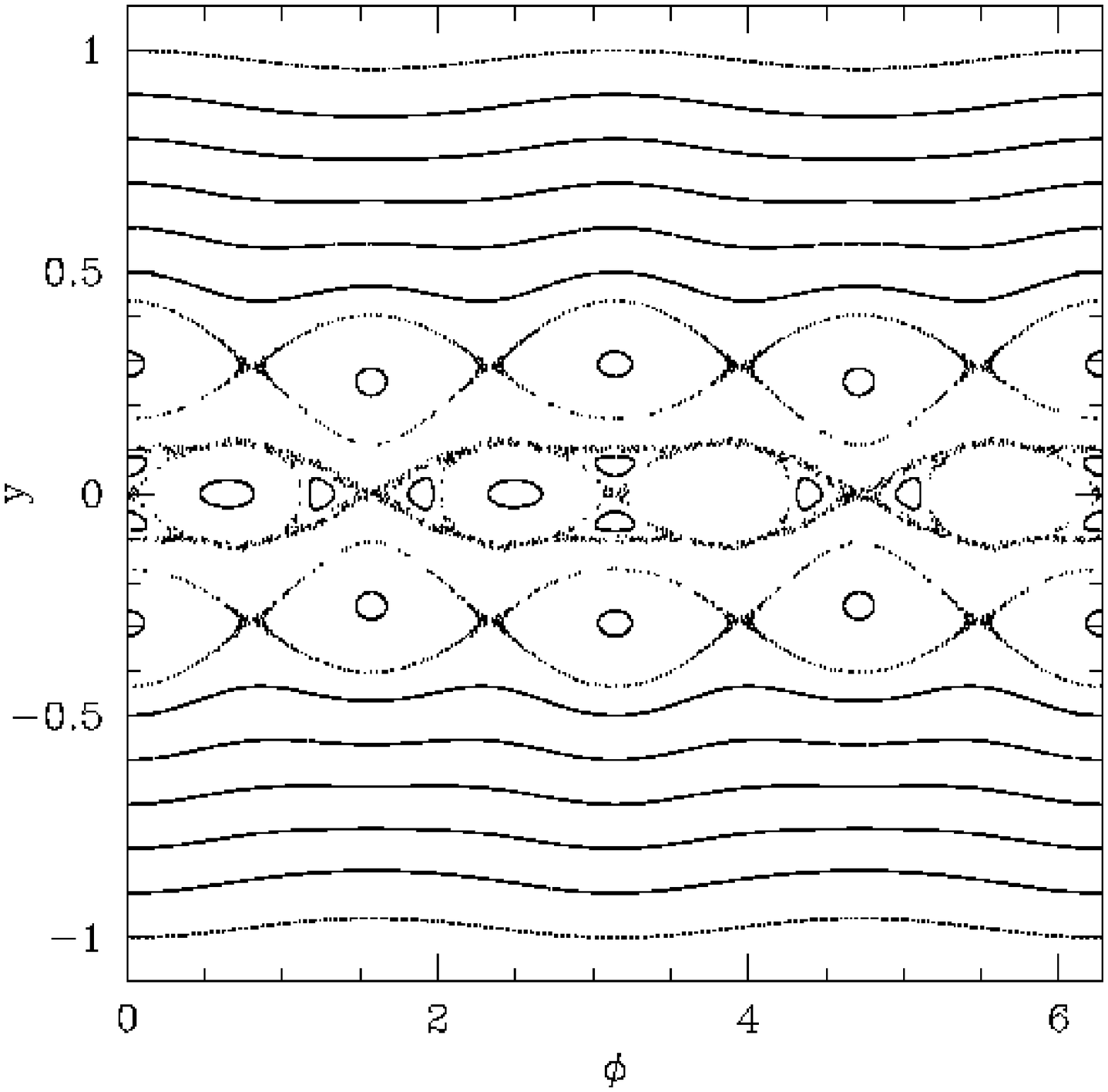}}}
\centerline{\hspace*{1.45in}$(c)$\hfill$(d)$\hspace{1.3in}}
\caption[Figure 9]{The map
(\ref{eq:mapone}--\ref{eq:mapthree}) with parameters $m=2$, $\epsilon=0.3$,
and (a) $\alpha=-0.25$; (b) $\alpha=-0.5$; (c) $\alpha=-0.75$. The orbits have
been iterated 1000 times, except for the chaotic orbits, which are
iterated 5000 times to provide better coverage.  (d) SOS for orbits in the
potential (\ref{eq:potdefnab}), with $\alpha=-0.5$ and $c=0.05$.  Each orbit
is plotted for 1000 periods, except for the chaotic orbit, which is plotted
for 10000 periods. }
\label{fi:conv}
\end{figure}

The maps with $m=2$, $\epsilon=0.3$ and $\alpha=-0.25$, $-0.5$,
and $-0.75$ are shown in Figure \ref{fi:conv}a,b,c. These show the
usual loop and box-like orbits, as well as a chaotic zone arising from a
single orbit. Panel (d) shows the SOS for orbits integrated in the potential
(\ref{eq:potdefnab}) with $\alpha=-0.5$, $c=0.05$, which shows good
qualitative agreement with the map in panel (b). 
Note that the stochastic web seen near $\alpha=0$ develops into a stochastic
sea by $\alpha=-0.75$. 

\def\lvec{{\bf L}}
\def\nvec{{\bf n}}\def\nlvec{{\bf l}}
\def\tvec{{\bf t}}

\section{Other potentials}\label{sec:other}

\subsection{$m=1$}

\begin{figure}
\centerline{\hbox{\epsfxsize=3in\epsfbox{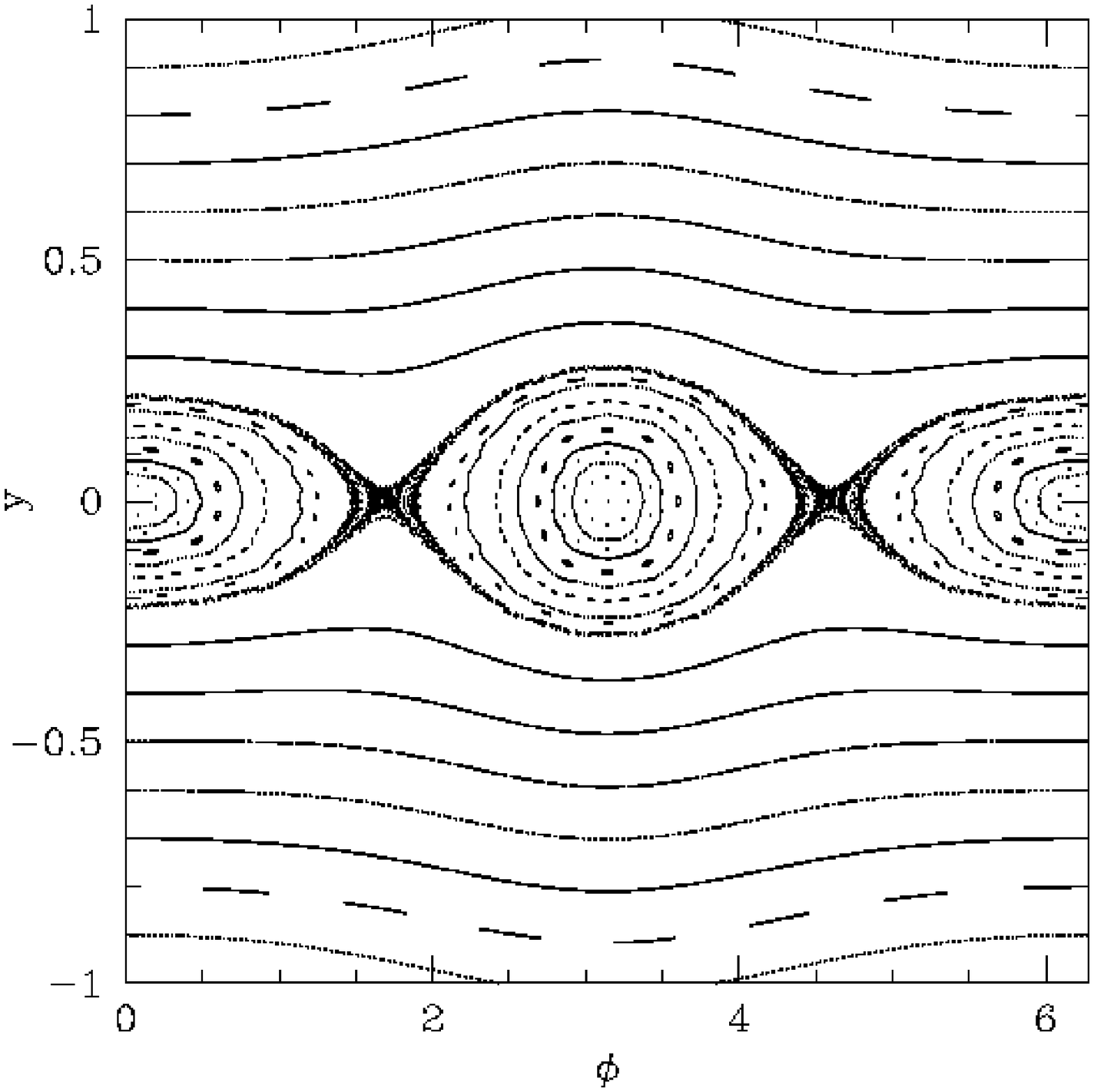}
                  \epsfxsize=3in\epsfbox{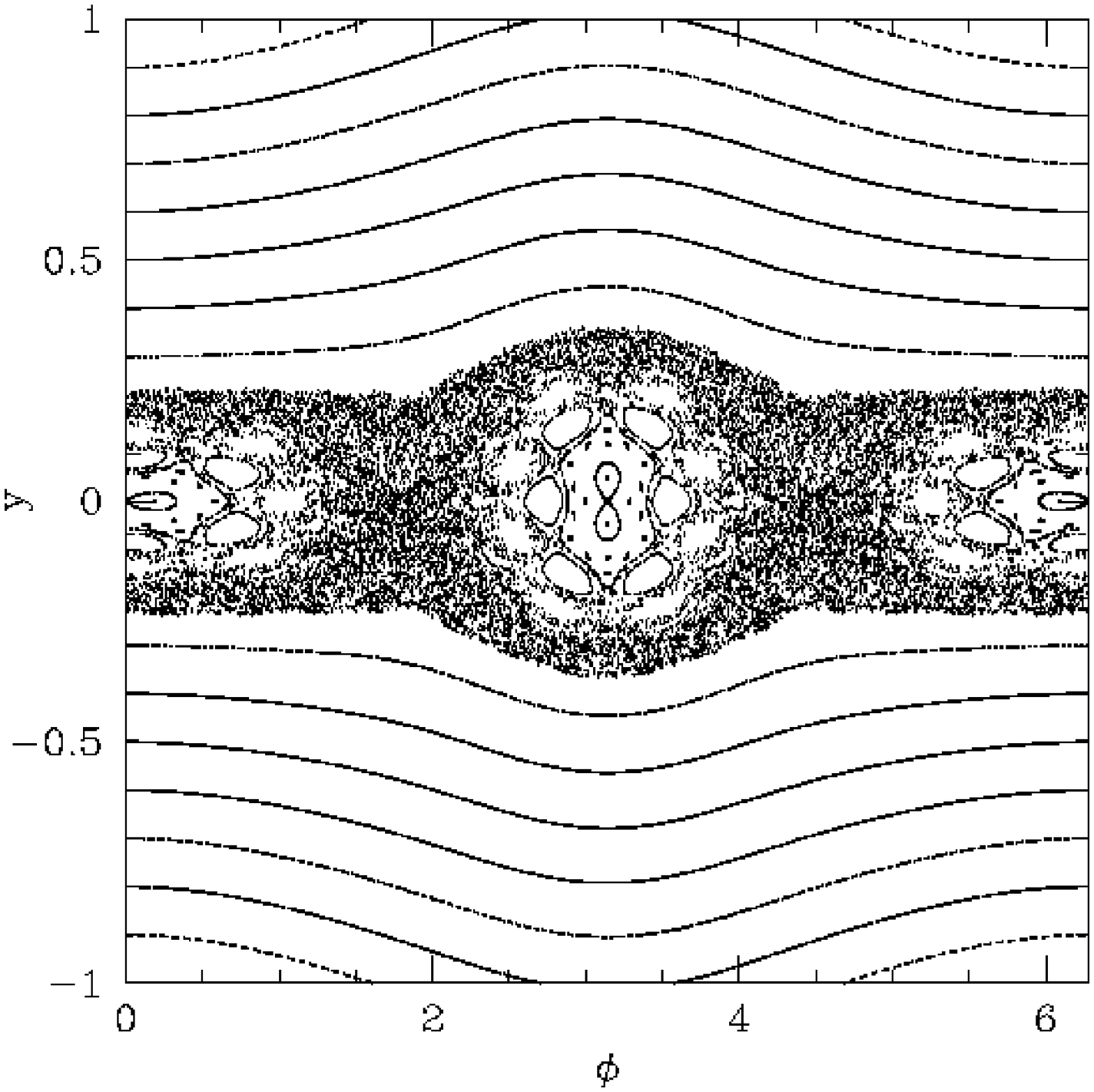}}}
\centerline{\hspace*{1.45in}$(a)$\hfill$(b)$\hspace{1.3in}}
\centerline{\hbox{\epsfxsize=3in\epsfbox{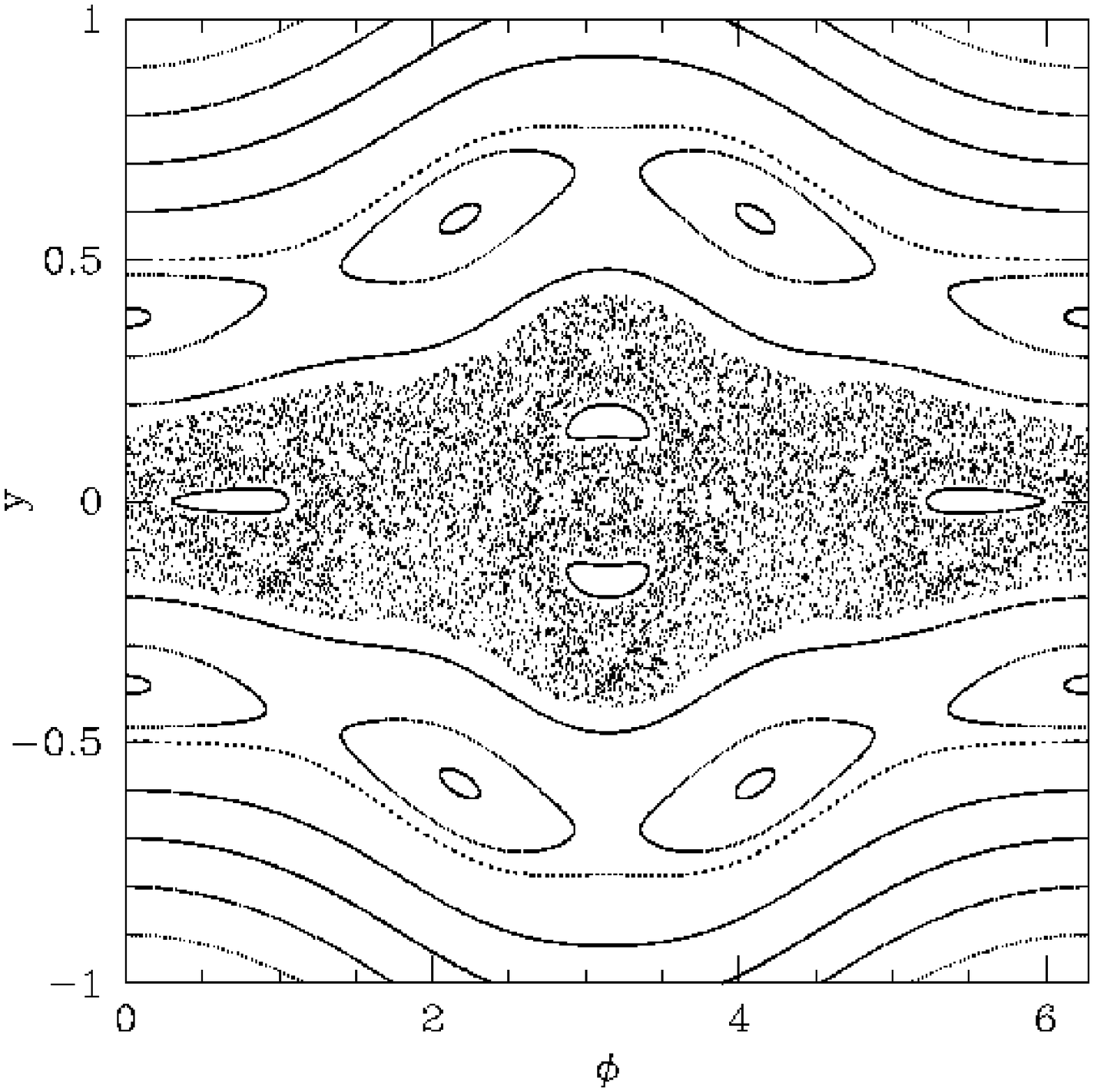}
                  \epsfxsize=3in\epsfbox{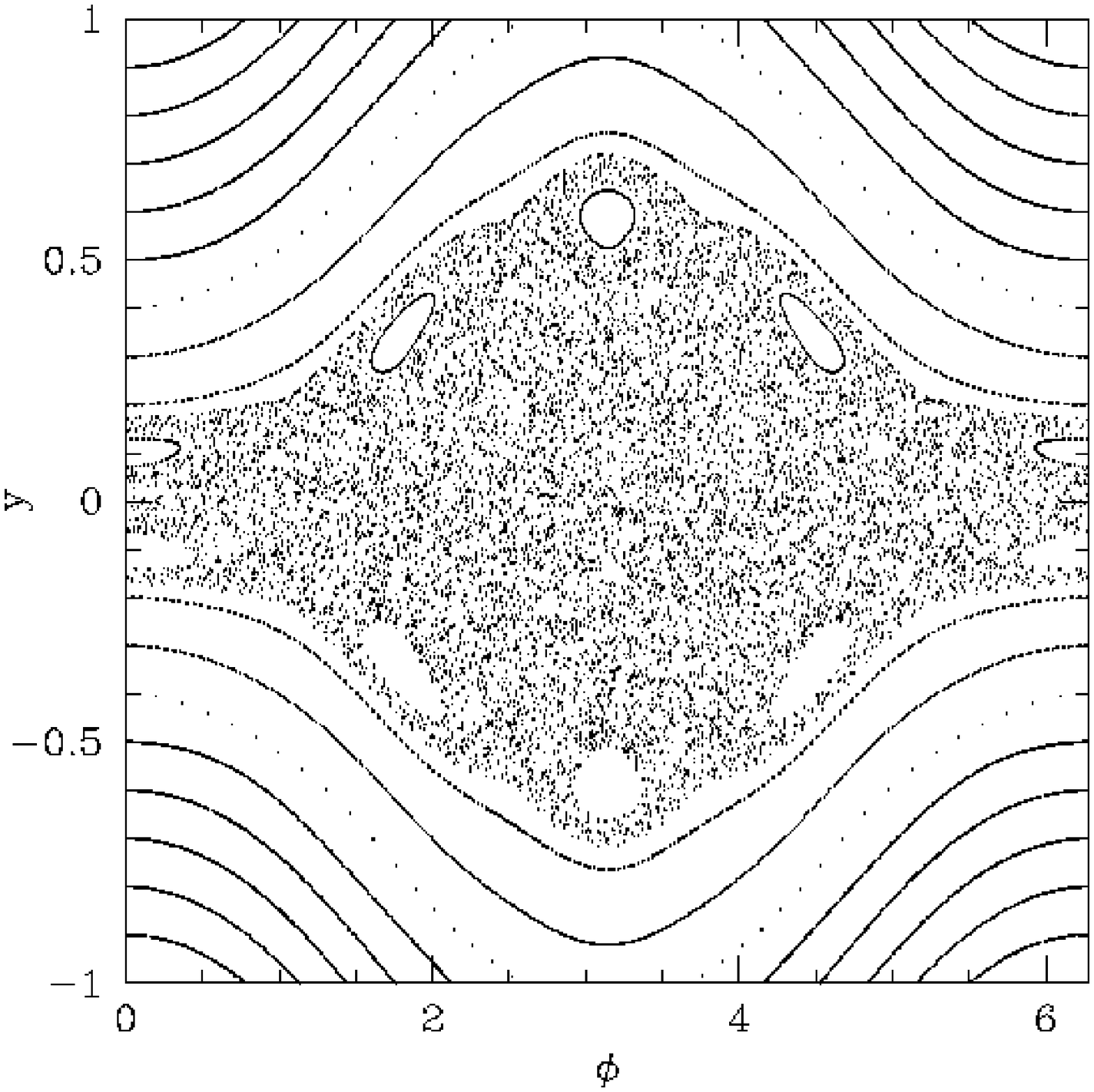}}}
\centerline{\hspace*{1.45in}$(c)$\hfill$(d)$\hspace{1.3in}}
\caption[Figure 10]{The map (\ref{eq:mapone}--\ref{eq:mapthree}) with
parameters $m=1$, $\epsilon=0.5$, and (a) $\alpha=1$; (b) $\alpha=0.5$; (c)
$\alpha=0$, (d) $\alpha=-0.5$. The orbits have been iterated 1000 times,
except for the chaotic orbits, which are iterated up to 20000
times to provide better coverage. The range $1<\alpha\le 2$ is not examined
because dipole potentials with this radial exponent cannot be generated by a
scale-free density distribution (cf. eq. \ref{eq:kkkk}).}
\label{fi:lop}
\end{figure}

We can use the map to explore the behaviour of orbits in lopsided ($m=1$)
potentials. In so doing we restrict ourselves to the range $\alpha\le1$
because dipole potentials with $1<\alpha\le 2$ cannot be generated by a
plausible density distribution (cf. eq. \ref{eq:kkkk}).
The maps with $m=1$, $\epsilon=0.5$ and $\alpha=1$, $0.5$, $0$, and $-0.5$ are
shown in Figure \ref{fi:lop}a,b,c,d. These show the usual loop orbits,
box-like orbits, and chaotic orbits.

The maps provide preliminary estimates of which orbits are needed to construct
self-consistent lopsided galaxies. The torques described by equation
(\ref{eq:mapone}) are generated by a density distribution that is elongated
along the axis $\phi=0$. When $\alpha=1$ (panel [a]) the loop orbits are most
elongated (smallest $|y|$) when their apocenters are near $\phi=1.8,4.6$ and
the box-like orbits have approximate $m=2$ symmetry around $\phi\simeq0,\pi$;
hence neither orbit family is likely to support a lopsided density
distribution. The situation is more promising for $\alpha\le0$ (panels [c] and
[d]). Here both the loop orbits and the chaotic sea are most elongated at
$\phi=0$. Thus it may be possible to construct self-consistent lopsided
galaxies with centrally concentrated density distributions.

\subsection{Potentials that are not scale-free}

In writing the precession rate as $g(\alpha,y)$ (eq. \ref{eq:gdef}), we have
assumed that the underlying axisymmetric potential is scale-free. More
generally the precession rate is a function of the energy of
the orbit in the underlying axisymmetric potential, $g(y,E)$, where $E=\half
v^2+\Phi(r)$ is assumed to be conserved since the non-axisymmetric component
of the potential is small. Apart from the energy-dependent precession rate the
map would remain unchanged. 

\subsection{Rotating potentials}

The map can also be used to study orbits in rotating potentials.
Let us suppose that the non-axisymmetric component of the potential is
stationary in a frame rotating with pattern speed $\Omega_p$, that the
potential has $m$-fold symmetry and that its azimuthal minimum lies along the
ray $\phi=\Omega_pt$. The azimuthal angle in the rotating frame may be written
$\psi\equiv \phi-\Omega_pt$. Then if $T_r(y,E)$ is the radial period of the
particle orbit in the underlying axisymmetric potential, the map 
(\ref{eq:mapone})--(\ref{eq:mapthree}) is easily generalized to
\begin{eqnarray}
y_n' & = & y_n-\half\epsilon\sin m\psi_n, \nonumber \\
\psi_{n+1}&=& \psi_n+g(y_n',E)-\Omega_p T_r(y_n',E),  \nonumber \\
y_{n+1} & = & y_n'-\half\epsilon\sin m\psi_{n+1}.
\end{eqnarray}
Note that $g$ is an odd function of $y$ while $T_r$ is even. 

\def\lvec{{\bf L}}
\def\yvec{{\bf Y}}
\def\nvec{{\bf n}}\def\nlvec{{\bf l}}
\def\tvec{{\bf t}}
\def\yhat{\hat{\bf y}}
\def\bnabla{\mbox{\boldmath $\nabla$}}

\subsection{Three dimensions}

We derive a mapping for motion in the three-dimensional generalization
of the potential (\ref{eq:potdefna}): 
\be
\Phi_\alpha({\bf r})=\left\{\begin{array}{cc}\hbox{sgn}(\alpha)
\left(x_1^2+{\textstyle x_2^2\over \textstyle b^2}+{\textstyle x_3^2\over 
\textstyle c^2}\right)^{\half\alpha}, 
&\alpha\not=0, \\ \half
\log\left(x_1^2+{\textstyle x_2^2\over \textstyle b^2}+{\textstyle x_3^2\over 
\textstyle c^2}\right), &\alpha=0. \end{array} \right.
\label{eq:3dpot}
\ee 

Here an orbit is defined by its vector angular momentum $\lvec$ and the unit
direction vector of its apocenter, $\nvec$, where $\lvec\cdot\nvec=0$. We
again prefer to deal with the dimensionless angular momentum, so we define
$\yvec = \lvec/L_c(E)$. The torque gives a kick to the angular momentum vector
which is concentrated at apocenter and equal to 
$-Q_{\alpha}{\nvec}_{n}\times{\tvec}_n$ where ${\tvec} = (n_x, n_y/b^2,
n_z/c^2)$. For nearly spherical potentials, 
$Q_{\alpha}$ is 
specified in (\ref{eq:delepsrat}). This kick is followed by a rotation of
$\nvec$ about the updated $\yvec$. Putting it all together,
\begin{eqnarray}
{\yvec}_n' & = & {\yvec}_n- \half Q_{\alpha}{\nvec}_{n} \times {\tvec}_n ,
\nonumber \\ 
{\nvec}_{n+1}&=& {\cal R}\big({\yvec}_n', g(\alpha, |{\yvec}_n'|)\big) \,
{\nvec}_{n}=-\frac{g(\alpha, |\yvec_{n}'|)}{|\yvec_{n}'|} 
\nvec_n \times \yvec_{n}', \nonumber  \\
{\yvec}_{n+1} & = & {\yvec}_n'-\half Q_{\alpha} {\nvec}_{n+1} \times
{\tvec}_{n+1},
\label{eq:threed}
\end{eqnarray}
where ${\cal R}({\bf v},\theta)$ is a rotation about axis ${\bf v}$ by
an angle $\theta$.

The mapping is apparently six-dimensional. However, it preserves the
unit magnitude of $\nvec$ and the orthogonality of $\yvec$ and
$\nvec$.  These two constraints on the motion reduce the dimension to
four. Also, the mapping is symplectic. The proof is deferred to
Appendix B.

A useful approximate description of the motion, valid when the potential is
not far from spherical,  is obtained by replacing the mapping equations
(\ref{eq:threed}) by the analogous differential equations
\begin{eqnarray}
{d\yvec\over dn} & = & -Q_{\alpha}{\nvec} \times {\tvec},\nonumber\\
{d\nvec\over dn} & = & -\frac{g(\alpha, |\yvec|)}{|\yvec|} \nvec \times \yvec,
\label{eq:eqmot}
\end{eqnarray}
where $n$ is the orbit number. 
These have the integral of motion (cf. eq. \ref{eq:appham}) 
\be
\overline H(\yvec,\nvec)=\int^yg(\alpha,y)dy+\half Q_\alpha\nvec\cdot\tvec.
\ee
where $y=|\yvec|$ and $\overline H$ is an averaged Hamiltonian analogous to
equation (\ref{eq:hamdefav}). This integral can be used to solve for $y$ in
terms of $\nvec$, so the equations of motion (\ref{eq:eqmot}) can be replaced
by equations for the two perpendicular unit vectors $\nvec$ and
$\yhat=\yvec/y$: 
\begin{eqnarray}
{d\yhat\over dn} & = & {Q_{\alpha}\over y}\left[\tvec\times \nvec-
\yhat \yhat\cdot(\tvec\times\nvec)\right],\nonumber\\
{d\nvec\over dn} & = & -g(\alpha,y)\nvec \times \yhat.
\label{eq:eqmota}
\end{eqnarray}
The solutions of these equations of motion can be investigated using a surface
of section (for example, defined by $\nvec\cdot\hat{\bf e}_2=0$), but we shall
not do so in this paper. 

\section{Discussion}

Our map provides a heuristic tool for studying the behaviour of eccentric
orbits in non-spherical potentials.  The map reproduces most of the
qualitative features of orbits in non-axisymmetric potentials similar to those
found in realistic galaxy models (e.g. Fig. \ref{fi:isomap}). The map can be
applied to a variety of potentials, as outlined in \S\ref{sec:other}. In a
limited sense, the map is more general than orbit integrations, since it does
not depend on the specific radial form of the non-axisymmetric potential so
long as the torque is concentrated near apocenter. The map is faster than
direct numerical integration by 2--3 orders of magnitude---even more for
near-radial orbits in centrally concentrated density distributions, which are
difficult to integrate accurately. The speedup offered by the map is
particularly important when exploring the long-term evolution of orbits in the
central regions of galaxies: at 10 pc from the center of a typical giant
galaxy the crossing time is only $10^{-5}$ of the Hubble time.

The map offers a powerful approach to studying many aspects of the behaviour
of orbits in triaxial potentials, including the distribution of Liapunov times
in chaotic orbits, the rate of mixing of a non-random distribution of stars,
trapping of chaotic orbits, and the long-term influence of a central black
hole on centrophilic orbits (Merritt 1996, Merritt \& Fridman 1996, 
Merritt \& Valluri 1996).

It may also be instructive to use the map to study the equilibrium and
stability of approximate stellar systems, in which the torque between two
elongated orbits is determined by the relative orientation of their
apocenters. A very similar approximation has been used as the basis of
heuristic explanations of bar formation in disk galaxies (Lynden-Bell 1979)
and the radial-orbit instability in spherical galaxies (Palmer \& Papaloizou
1987).

\begin{acknowledgements}

We thank Jeremy Goodman for thoughtful advice. This research was
supported in part by NSERC. JT acknowledges the support of 
the Harlan Smith Fellowship under NASA grant NAGW 1477, and ST acknowledges
the support of an Imasco Fellowship.

\end{acknowledgements}

\appendix

\section{Power series for the precession rate in power-law 
potentials}\label{sec:appa1}

The goal of this Appendix is to provide a power-series expansion for the
precession rate $g(\alpha,y)$ (eq. \ref{eq:gdef}). For the sake of brevity we
focus first on potentials with $\alpha>0$ (cf. eq. \ref{eq:potdef}), extending
the results to $\alpha\le0$ at the end. 

We first extend the domain of $g$ by defining $g(\alpha,y)=0$ for $y>1$. We
then take the Mellin transform
\be
G(\alpha,s)\equiv \int_0^\infty g(\alpha,y)y^{s-1}dy;
\ee
upon evaluating the integrals we have
\be
G(\alpha,s)={\pi^{1/2}2^{s/2}\over\alpha h^s(\alpha)}
{\Gamma(\half+\half s)\Gamma(s/\alpha)
\over\Gamma(1+\half s+s/\alpha)},\qquad \alpha>0,
\label{eq:gggdef}
\ee
which holds in the right half-plane Re$(s)>0$. 

The inverse Mellin transform is 
\be 
g(\alpha,y)={1\over2\pi i}\int_{\sigma-i\infty}^{\sigma+i\infty}G(\alpha,s)
y^{-s}ds,
\label{eq:invmel}
\ee 
where $\sigma$ is real and positive. For $0<y<1$, the integration contour
can be closed by a semi-circle in the negative half-plane from $\sigma+i\infty$
to $\sigma-i\infty$. Since $1/\Gamma(z)$ is entire, the only contributions to
the contour integral (\ref{eq:invmel}) 
arise from poles in the functions $\Gamma(z)$ in the numerators of equations
(\ref{eq:gggdef}), which occur at $z=-n$, $n=0,1,2,\ldots$ and have residues
$(-1)^n/\Gamma(n+1)$. Thus
\begin{eqnarray}
g(\alpha,y) & = & \pi^{1/2}\sum_{n=0}^\infty {(-1)^n h^{2n+1}(\alpha)
\Gamma[-(2n+1)/\alpha] \over 
2^{n-1/2}\alpha\Gamma[\half-n-(2n+1)/\alpha]\Gamma(n+1)}y^{2n+1} \nonumber \\
&{}& +
\pi^{1/2}\sum_{n=0}^\infty {(-1)^nh^{\alpha n}(\alpha)\Gamma[(1-\alpha n)/2]
\over 2^{\alpha n/2}\Gamma(1-n-\alpha n/2)\Gamma(n+1)}y^{\alpha n}.
\end{eqnarray}
Using the identity $\Gamma(z)\Gamma(1-z)=\pi/\sin\pi z$ this expression can be
rewritten as
\begin{eqnarray}
g(\alpha,y) & = & 
\pi-\pi^{1/2}\sum_{n=0}^\infty { h^{2n+1}(\alpha)\cot[\pi(2n+1)/\alpha]
\Gamma[\half+n+(2n+1)/\alpha]
\over 2^{n-1/2}\alpha\Gamma[1+(2n+1)/\alpha]\Gamma(n+1)}y^{2n+1} \nonumber \\
& {} & + \pi^{1/2}
\sum_{n=1}^\infty {h^{\alpha n}(\alpha)\tan(\half\pi\alpha n)
\Gamma[n(1+\alpha/2)]
\over 2^{\alpha n/2}\Gamma[(1+\alpha n)/2]\Gamma(n+1)}
y^{\alpha n},\qquad \alpha>0.
\label{eq:asymp}
\end{eqnarray}
In the limit $y\to 0$ we recover equations (\ref{eq:zerolim}) and
(\ref{eq:gasymp}). 

For $\alpha<0$ we can apply the relation (\ref{eq:dual}) to obtain
\begin{eqnarray}
g(\alpha,y) & = & 
{2\pi\over2-\absa}
+\pi^{1/2}\sum_{n=0}^\infty {h^{2n+1}[2\absa/(2-\absa)]\tan[\pi(2n+1)/\absa]
\Gamma[(2n+1)/\absa]\over 2^{n-1/2}\absa
\Gamma[\half-n+(2n+1)/\absa]\Gamma(n+1)}y^{2n+1} \nonumber \\
& {} & +\pi^{1/2}
\sum_{n=1}^\infty { h^{2\absa n/(2-\absa)}[2\absa/(2-\absa)]
\tan[\pi\absa n/(2-\absa)]\Gamma[2n/(2-\absa)]\over 
2^{\absa n/(2-\absa)-1}(2-\absa)\Gamma[\half+\absa n/(2-\absa)]\Gamma(n+1)}
y^{2\absa n/(2-\absa)},
\label{eq:asympone}
\end{eqnarray}
where $\beta=-\alpha$, $\alpha<0$. 
The series (\ref{eq:asymp}) and (\ref{eq:asympone}) are asymptotic: if
truncated after a fixed number of terms the series become arbitrarily
accurate as $y\to0$, but for fixed $y$ the series do not generally converge
as more terms are added.

\section{The three-dimensional mapping is symplectic}\label{sec:appa2}

We claim that the generalization of the two-dimensional mapping
(eqs. \ref{eq:mapone}--\ref{eq:mapthree}) to triaxial potentials
(eqs. \ref{eq:threed}) is symplectic. There are two routes to the proof: one
familiar but messy, the other less familiar but more elegant and
illuminating. We outline the first and describe the second in detail. On the
messy road, we define two pairs of canonically conjugate variables.  One pair
joins the magnitude of the angular momentum and the argument of apocentre and
the other the $x_3$-component of the angular momentum vector and the longitude
of the ascending node on the $x_1$-$x_2$ plane. That these coordinates are
canonically conjugate is a basic result of classical celestial mechanics
(e.g. Plummer 1960). Then we express the map in terms of these coordinates,
derive the Jacobian ${\bf M}$ of the mapping transformation, and show after
some algebraic labor, that this Jacobian fulfills the requirements that a
canonical transformation must satisfy, namely: ${\bf M}^{T}{\bf J}{\bf M} =
{\bf J}$, where ${\bf J}$ is the $4\times4$ canonical symplectic matrix (see
Goldstein 1980 for definitions and details).

We prefer to observe that the mapping derives from the Hamiltonian:
\be
H = H_y(\yvec) + {\delta}_{1}(t) H_n(\nvec),
\label{eq:appham} 
\ee
where $H_y=\int^yg(\alpha,y)dy$, $y=|\yvec|$, and
$H_n=\half Q_{\alpha} \nvec \cdot \tvec$, with the help of the bracket:
\be
\{ F_1(\yvec, \nvec), F_2(\yvec, \nvec) \} = \yvec \cdot ({\bnabla}_{\yvec}F_1
\times {\bnabla}_{\yvec}F_2) 
+\nvec \cdot ({\bnabla}_{\nvec}F_1 \times {\bnabla}_{\yvec}F_2 +
{\bnabla}_{\yvec}F_1 \times {\bnabla}_{\nvec}F_2). 
\ee
The bracket
\footnote{As far as $H_y$ and $H_n$ are concerned, the first
term in the bracket does not contribute anything. But, it could
certainly come alive when examining orbits in a rotating potential.}
gives the time evolution of any real valued function $F_1$ of these
two vectors, along the vector field generated by another such function
$F_2$, through the relation
\be
{d\over dt}F_1(\yvec,\nvec)=\{F_1,F_2\}.
\ee

The equations for the time evolution 
of $\yvec$ and $\nvec$ along the vector field generated by $H_y$ are: 
\begin{eqnarray}
\frac{d}{dt}\yvec & = & 0, \nonumber \\
\frac{d}{dt}\nvec & = & - \frac{g(\alpha,|\yvec|)}{|\yvec|} \nvec \times \yvec.
\end{eqnarray}
The first of these equations states that $\yvec$ is conserved by 
$H_y$. The second is telling us $H_y$ causes a rotation of 
$\nvec$ by angle $g$ about vector $\yvec$. This is one of the mapping 
steps. $H_n$ on the other hand generates motion: 
\begin{eqnarray}
\frac{d}{dt}\yvec & = & -Q_{\alpha} \nvec \times \tvec, \nonumber \\
\frac{d}{dt}\nvec & = & 0.
\end{eqnarray}
When we compose the action of $H_y$ and $H_n$, in the manner dictated
by the periodic delta function, we recover the mapping (\ref{eq:threed}).
Now we still have
to show that the dynamics generated by this bracket is canonical. The
bracket is known as a Lie-Poisson bracket and is closely related to
the bracket used in the study of conservative rigid body dynamics (see
Marsden 1992 for a general discussion, and Touma and Wisdom 1994 for a
discussion of Lie-Poisson brackets in an astronomical context). On a
suitably defined pair of canonical variables, the bracket reduces to
the canonical Poisson bracket and the dynamics generated by the
bracket consists of canonical transformations. Thus, $H_y$ and $H_n$
separately generate symplectic mappings and their composition via the
mapping is therefore symplectic, completing our proof.


\begin{thebibliography}{}

\bibitem[Binney (1982)]{bin82}Binney, J. 1982, MNRAS 201, 1

\bibitem[Binney \& Spergel (1982)]{binspe82}Binney, J., \& Spergel, D. 1982,
ApJ 252, 308

\bibitem[Binney \& Tremaine (1987)]{bintre87}Binney, J. J., \& Tremaine,
S. 1987, Galactic Dynamics (Princeton: Princeton University Press)

\bibitem[de Zeeuw (1985)]{dez85}de Zeeuw, T. 1985, MNRAS 216, 273

\bibitem[Gebhardt et al. (1996)]{geb96}Gebhardt, K., Richstone, D., Ajhar,
E. A., Lauer, T. R., Byun, Y.-I., Kormendy, J., Dressler, A., Faber, S. M.,
Grillmair, C., and Tremaine, S. 1996, AJ 112, 105

\bibitem[Gerhard \& Binney (1985)]{gerbin85}Gerhard, O. E., \& Binney,
J. 1985, MNRAS 216, 467

\bibitem[Goldstein (1980)]{gold80}Goldstein, H., 1980, Classical Mechanics, 
2nd ed. (Reading: Addison-Wesley)

\bibitem[Grant \& Rosner (1994)]{graros94}Grant, A. K., \& Rosner, J. L. 1994,
Am. J. Phys. 62, 310

\bibitem[Kormendy \& Richstone (1995)]{korric95}Kormendy, J., \& Richstone,
D. 1995, ARAA 33, 581

\bibitem[Kuijken (1993)]{kui93}Kuijken, K. 1993, ApJ 409, 68

\bibitem[Lees \& Schwarzschild (1992)]{leesch92}Lees, J. F., \& Schwarzschild,
M. 1992, ApJ 384, 491

\bibitem[Lichtenberg \& Lieberman (1992)]{liclie92}Lichtenberg, A. J., 
\& Lieberman, M. A., 1992, Regular and Chaotic Dynamics, 2nd ed.  
(Springer, New York)

\bibitem[Lynden-Bell (1962)]{dlb62}Lynden-Bell, D. 1962, MNRAS 124, 95

\bibitem[Lynden-Bell (1979)]{dlb79}Lynden-Bell, D. 1979, MNRAS 187, 101

\bibitem[Marsden (1992)]{mars92}Marsden, J. E. 1992, Lectures on Mechanics
(Cambridge: Cambridge University Press)

\bibitem[Merritt (1996)]{mer96}Merritt, D. 1996, Science 271, 337

\bibitem[Merritt \& Fridman (1996)]{merfril96}Merritt, D., \& Fridman,
T. 1996, ApJ 460, 136

\bibitem[Merritt \& Valluri (1996)]{merval96}Merritt, D., \& Valluri, M. 1996,
ApJ 471, 82

\bibitem[Miralda-Escud\'e \& Schwarzschild (1989)]{mirsch89}Miralda-Escud\'e,
J., \& Schwarzschild, M. 1989, ApJ 339, 752

\bibitem[Palmer \& Papaloizou (1987)]{palpap87}Palmer, P. L., \& Papaloizou,
J. 1987, MNRAS 224, 1043

\bibitem[Pfenniger \& de Zeeuw (1989)]{pfedez88}Pfenniger, D., \& de Zeeuw,
T. 1989, in Dynamics of Dense Stellar Systems, ed. D. Merritt (Cambridge:
Cambridge University Press), 81

\bibitem[Plummer (1960)]{plum60}Plummer, H. C. 1960, An Introductory 
Treatise on Dynamical Astronomy (New York: Dover)

\bibitem[Richstone (1980)]{ric80}Richstone, D. O. 1980, ApJ 238, 103

\bibitem[Richstone (1982)]{ric82}Richstone, D. O. 1982, ApJ 252, 496

\bibitem[Schwarzschild (1979)]{sch79}Schwarzschild, M. 1979, ApJ 232, 236

\bibitem[Schwarzschild (1993)]{sch93}Schwarzschild, M. 1993, ApJ 409, 563

\bibitem[Sridhar \& Touma (1997)]{sritou97} Sridhar, S., \& Touma, J. 1997,
MNRAS 287, L1

\bibitem[Touma \& Wisdom (1994)]{touwis96} Touma, J., \& Wisdom, J. 1994, 
AJ 107, 1189

\bibitem[Whittaker (1959)]{whi59}Whittaker, E. T. 1959, A Treatise on the
Analytical Dynamics of Particles and Rigid Bodies, 4th ed. (Cambridge:
Cambridge University Press)

\end{thebibliography}
\end{document}